\pdfoutput=1
\documentclass[11pt]{article}
\usepackage{amsfonts}
\usepackage{amsmath,amssymb,amsthm}
\usepackage{mathrsfs}
\usepackage{float}

\usepackage[utf8]{inputenc}
\usepackage{caption,xspace}
\usepackage{subcaption}
\numberwithin{equation}{section}

\usepackage{graphicx}
\newcommand{\FMDG}[1]{\includegraphics{#1}}

\usepackage[sort&compress,numbers,colon,merge]{natbib}
\usepackage{multirow}
\usepackage{color}
\usepackage[colorlinks]{hyperref}
\hypersetup{
pdfauthor={Tony Gherghetta, Benedict von Harling, Anibal D. Medina and Michael A. Schmidt},
pdftitle={The Scale-Invariant NMSSM and the 126 GeV Higgs Boson}
}

\usepackage{verbatim}
\usepackage{float}
\usepackage{tikz}
\usetikzlibrary{shapes,arrows}
\newcommand{\be}{\begin{equation}}
\newcommand{\ee}{\end{equation}}
\newcommand{\bea}{\begin{eqnarray}}
\newcommand{\eea}{\end{eqnarray}}

\newcommand{\met}{\ensuremath{\not\!\!E_T}\xspace}
\def\4vol{{\int d^4x \sqrt{-g}}}

\def\beq{\begin{equation}}
\def\eeq{\end{equation}}
\def\bea{\begin{eqnarray}}
\def\eea{\end{eqnarray}}
\def\bitem{\begin{itemize}}
\def\eitem{\end{itemize}}

\newcommand{\footlabel}[2]{%
    \addtocounter{footnote}{1}%
    \footnotetext[\thefootnote]{%
        \addtocounter{footnote}{-1}%
        \refstepcounter{footnote}\label{#1}%
        #2%
    }%
    $^{\ref{#1}}$%
}

\newcommand{\footref}[1]{%
    $^{\ref{#1}}$%
}

\newcommand{\Eqref}[1]{Eq.~(\ref{#1})}

\newcommand{\Figref}[1]{Fig.~\ref{#1}}
\newcommand{\Tabref}[1]{Tab.~\ref{#1}}
\newcommand{\Secref}[1]{sec.~\ref{#1}}

\newcommand{\eV}{\ensuremath{\,\mathrm{eV}}}

\newcommand{\GeV}{\ensuremath{\,\mathrm{GeV}}}
\newcommand{\TeV}{\ensuremath{\,\mathrm{TeV}}}

\newcommand{\Mmess}{\Lambda_{\rm mess}}
\newcommand{\msoft}{m_{\rm soft}}

\newcommand{\soft}{{\rm soft}}

\newcommand{\nc}{\newcommand}

\nc{\nt}{\tilde{N}}
\nc{\ra}{\rightarrow}
\nc{\lsim}{\begin{array}{c}\,\sim\vspace{-21pt}\\< \end{array}}
\nc{\gsim}{\begin{array}{c}\sim\vspace{-21pt}\\> \end{array}}
\nc{\tnt}{\tilde{N}}
\nc{\tst}{\tilde{t}}

\nc{\LL}{L}
\nc{\vv}{\tilde{v}}

\title{}

\begin{document}
\allowdisplaybreaks[1]

\begin{titlepage}

\begin{center}
{\Large\textbf{The Scale-Invariant NMSSM and the 126 GeV Higgs Boson}}
\\[10mm]
{\large
Tony Gherghetta$^{a,b,}$\footnote{\texttt{tgher@unimelb.edu.au}},
Benedict von Harling$^{a,c,}$\footnote{\texttt{bharling@sissa.it}},
Anibal D. Medina$^{a,}$\footnote{\texttt{anibal.medina@unimelb.edu.au}},
Michael A.~Schmidt$^{a,}$\footnote{\texttt{michael.schmidt@unimelb.edu.au}}}
\\[5mm]
{\small\textit{$^a$ARC Centre of Excellence for Particle Physics at the Terascale,\\
School of Physics, The University of Melbourne, Victoria 3010, Australia\\
$^b$Stanford Institute of Theoretical Physics, Stanford University, Stanford, CA 94305, USA\\
$^c$SISSA, Via Bonomea 265, 34136 Trieste, Italy}
}
\end{center}

\vspace*{1.0cm}
\date{\today}

\begin{abstract}
\noindent
The recent LHC discovery of a Higgs-like resonance at 126 GeV suggests that the minimal supersymmetric 
standard model must be modified in order to preserve naturalness. A simple extension is to include a singlet 
superfield and consider the scale-invariant NMSSM, whose renormalizable superpotential contains no 
dimensionful parameters. This extension not only solves the $\mu$-problem, but can easily accommodate a 
126 GeV Higgs. We study the naturalness of the scale-invariant NMSSM taking into account the recent constraints 
from LHC searches, flavor physics and electroweak precision tests. We show that  TeV-scale stop masses are still allowed in much of the parameter space with 5$\%$ tuning for a low messenger scale of 20 TeV, split families (with 
third-generation sleptons decoupled) and Higgs-singlet coupling $\lambda$ of order one. For larger values of the Higgs-singlet coupling, which can relieve 
the tuning in the Higgs VEV, an additional tuning in the Higgs mass limits increasing the (lightest)
stop mass beyond 1.2 TeV, the gluino mass above 3 TeV, and electroweak charginos and neutralinos beyond 
400 GeV for a combined tuning better than $5\%$. This implies that the natural region of parameter space for the 
scale-invariant NMSSM will be fully explored at the 14 TeV LHC.

\end{abstract}

\end{titlepage}

\setcounter{footnote}{0}



\section{Introduction}

The recent discovery of a resonance with Higgs-like properties and mass around 126 GeV at the Large Hadron Collider (LHC) provides compelling evidence that the Higgs 
mechanism is responsible for the generation of elementary particle masses in the Standard Model (SM). An important next question is to ascertain the 
nature of  this new resonance by measuring its properties and couplings.  Preliminary indications already suggest an enhancement of the Higgs 
decay rate to photons which would signal the presence of new light states beyond the standard model. A well-motivated possibility is to identify the 
new resonance with the lightest Higgs boson in the supersymmetric standard model. Supersymmetry (SUSY) provides a natural solution to the hierarchy problem 
while simultaneously allowing for gauge coupling unification and a suitable dark matter candidate~\cite{Martin:1997ns}. However, a Higgs boson at 126 GeV is in extreme tension with maintaining naturalness in the Minimal Supersymmetric Standard Model (MSSM).  As is well-known, large radiative corrections from stop loops 
are required to enhance the Higgs quartic coupling which originates at tree-level from $D$-terms. This is accomplished by either having large ($\gg 1$ TeV) stop 
masses, in the case of negligible mixing in the stop sector, or a large trilinear $A_t$-term. In either case, a fine-tuned cancellation caused by this little hierarchy 
is needed to obtain the correct electroweak symmetry breaking scale.

To alleviate the dependence on large stop masses or the $A_t$-term in obtaining a sufficiently large Higgs quartic coupling, the MSSM can be 
extended to include a gauge-singlet chiral superfield $S$ in addition to the MSSM particle content. In this Next-to-Minimal Supersymmetric Standard Model (NMSSM; see~\cite{Ellwanger:2009dp} for a review) the singlet superfield couples to the up and down Higgs superfields, $H_u$ and $H_d$, via the superpotential term $\lambda S H_u H_d$, where 
$\lambda$ is a dimensionless coupling. This modifies the upper bound on the Higgs mass at tree-level with respect to the MSSM such that
\begin{equation}
m_h^2\leq m_Z^2 \cos^2 2\beta +\lambda^2 v^2\sin^2 2\beta,
\label{HiggsMassBoundIntro}
\end{equation}
where $v$ is the electroweak vacuum expectation value (VEV) and $\tan\beta$ denotes the ratio of the up and down Higgs VEVs.  The extra tree-level contribution in~\eqref{HiggsMassBoundIntro} from the singlet coupling $\lambda$ is maximal for $\tan\beta\sim 1$ and can thus increase the Higgs mass, thereby alleviating the need for large stop masses or a large $A_t$-term and the resulting fine-tuning problem.  In addition, values of $\lambda\gtrsim 1$ further ameliorate the fine-tuning problem by suppressing the sensitivity of the Higgs VEV with respect to changes in the soft masses~\cite{Barbieri:2006bg,Hall:2011aa}. 

In addition to the Higgs-singlet coupling there can also be other singlet field couplings, including explicit mass terms in the superpotential. These mass terms can be avoided by considering the scale-invariant version of the NMSSM which is characterized by forbidding all dimensionful terms in the superpotential at the renormalizable level. 
This can be accomplished via a discrete $\mathbb{Z}_3$ symmetry under which all superfields are charged, only allowing the superpotential terms $\lambda S H_u H_d+(\kappa/3) S^3$ (the resulting domain wall problem in the early universe can be addressed as discussed e.g.~in \cite{Abel:1996cr,Panagiotakopoulos:1998yw}). This scenario has the additional virtue of providing a solution to the $\mu$-problem encountered in the MSSM once the scalar component of $S$ acquires a VEV of order the electroweak scale.

In this paper, we  study the parameter space of the scale-invariant NMSSM. We search for regions that can incorporate a Higgs at 126 GeV and that are in accordance with all experimental constraints. Our main focus are the constraints on the sparticle spectrum coming from naturalness. To quantify the naturalness, we use the usual logarithmic-derivative measure of fine-tuning. We not only include the one-loop radiative corrections coming from supersymmetric running in the calculation of the fine-tuning measure but also contributions from the Coleman-Weinberg potential, which are particularly important for (s)particles coupling strongly to the Higgs-singlet sector.
In our analysis we will make a number of assumptions.
First, the only superpartners that we consider in addition to those in the Higgs-singlet sector are stops, sbottoms and gauginos. All other sfermions are taken heavy enough to be decoupled from our effective theory.  Decoupling these sfermions, corresponding to split families (including the third generation sleptons), helps to satisfy several collider and flavor constraints but does not affect the naturalness of the Higgs sector because these sparticles couple only weakly to the Higgs. 
Furthermore, in order to allow for a large value of $\lambda$ and to minimize loop corrections to the soft parameters, we consider a low cutoff of our effective theory, $\Mmess\simeq 20$--$1000$ TeV.
This scale could be associated with a messenger sector like in gauge mediation. Alternatively, it could correspond to a cutoff below which the Higgs sector emerges as composites of an underlying strongly coupled theory as in accidental SUSY~\cite{Gherghetta:2003he,Sundrum:2009gv,Gherghetta:2011wc} or fat Higgs models~\cite{Harnik:2003rs}. 
Finally, we allow all values for $\lambda$ and $\kappa$ that are consistent with perturbativity up to the cutoff scale $\Mmess$. This covers not only the parameter space of the usual NMSSM (for which one typically requires perturbativity of the couplings up to the GUT scale)  but also includes values $\lambda \gtrsim 0.7$ that develop a Landau pole below the GUT scale as considered in $\lambda$SUSY models~\cite{Barbieri:2006bg}. As shown recently~\cite{Hardy:2012ef}, a Landau pole in the coupling $\lambda$ is not necessarily detrimental to maintaining gauge coupling unification. 

The assumptions that we make -- split families (with third-generation sleptons decoupled), a low cutoff $\Mmess$ and a possibly large coupling $\lambda$ -- correspond to an optimal-case scenario for the (scale-invariant) NMSSM from the viewpoint of naturalness. Indeed, if the families are not split, stops and sbottoms would tend to be much heavier due to the strong collider constraints on first-generation squarks. Similarly, a larger cutoff scale $\Mmess$ would result in larger loop corrections to soft parameters from supersymmetric running. Large values $\lambda \gtrsim 0.7$, on the other hand, alleviate the fine-tuning problem as discussed below Eq.~\eqref{HiggsMassBoundIntro}. This ensures that our constraints from naturalness are conservative in the sense that a different implementation of the (scale-invariant) NMSSM would have a smaller region of parameter space for a given amount of fine-tuning.

The amelioration of the Higgs VEV tuning at large $\lambda$ seems to allow for stop masses above the TeV scale for the same level of tuning as in the MSSM. However, we point out that there is also a competing effect. The bound in Eq.~\eqref{HiggsMassBoundIntro} implies that the Higgs mass at tree-level grows with $\lambda$ if $\tan \beta$ is small. Electroweak precision tests, on the other hand, require that $\tan \beta \sim 1$ for large $\lambda$. This means that the Higgs mass typically becomes larger than ${126 \GeV}$ in the region $\lambda \gtrsim 1$. A tuning in the various contributions to the Higgs mass is then required to bring the Higgs mass down to ${126 \GeV}$. This additional tuning prevents the stop masses to be raised much above the TeV scale if the theory is supposed to remain natural. The two competing effects can be incorporated into a single measure of tuning, which we define as the product of the Higgs VEV and Higgs mass tunings. We find that this combined tuning measure is minimized for $\lambda\approx 1$, leading to restrictions on a natural sparticle spectrum.

We use the numerical package {\it NMSSMTools} 3.2.1 to compute the sparticle spectrum, but modify the program to allow for a small messenger scale $\Mmess$ and a large  coupling $\lambda$. We impose constraints from direct searches at colliders and flavor physics. Several such constraints are already implemented in {\it NMSSMTools}. We complement these with the recent limits on heavy $CP$-even Higgses from ATLAS and CMS, as well as limits from stop and gluino searches. In addition, we require that the Peskin-Takeuchi parameters $S$ and $T$ are consistent with their measured ranges. Finally, we verify that the LSP relic abundance does not overclose the universe and impose constraints from direct detection. 
We then perform a numerical scan of the parameter space of the scale-invariant NMSSM which satisfies these constraints and optimize the scan for regions that have a small tuning in the Higgs VEV and a Higgs mass near $126 \GeV$. 
The results from the scan leads to restrictions on a natural sparticle spectrum. It is found that for a total tuning which is better than $5\%$, and $\Mmess\simeq 20$ TeV, the lightest stop must have a mass below 1.2 TeV, the gluino should be below 3 TeV, and the electroweak charginos and neutralinos must lie below 400 GeV. These mass ranges still leave room for naturalness to be explored by the 14 TeV LHC. Furthermore if the total tuning is relaxed to be better than $1\%$ then are a few points where a large enhancement in the Higgs diphoton signal strength can be obtained.

There have been a number of previous analyses~\cite{Barbieri:2006bg,Hall:2011aa,Franceschini:2010qz} where non-minimal supersymmetry with large $\lambda$ was studied in the context of naturalness. These differ from our work in that they do not perform a full scan of the parameter space. Furthermore, Refs.~\cite{Barbieri:2006bg,Hall:2011aa} include explicit mass terms (such as a $\mu$-term and a singlet  supersymmetric mass) in the superpotential. These papers also neglect the quantum effects accounted for by the Coleman-Weinberg potential which we find to be important. Naturalness and a 126 GeV Higgs boson in the context of the NMSSM
with perturbativity up to the GUT scale (i.e. $\lambda \lesssim 0.7$) was considered e.g.~in~\cite{Kang:2012sy,King:2012is,*King:2012tr,Vasquez:2012hn,Cao:2012fz,*Cao:2012yn}. Our work also complements the recent analysis in Ref.~\cite{Agashe:2012zq}, where stop masses were restricted to lie below 500 GeV.

The paper is organized as follows. In \Secref{sec:NMSSM}, we present the Higgs sector of the scale-invariant NMSSM and include the contributions from the effective potential. A brief description of the numerical scan with the range of the model parameters is given in \Secref{sec:NumAnal}. The naturalness constraints are discussed in \Secref{naturalness}, where fine-tuning can arise from both the electroweak VEV and the Higgs mass. Requiring small combined tuning leads to constraints on the stop masses and the Higgs-singlet coupling $\lambda$. Further constraints are described in \Secref{sec:collider}, which arise
from collider limits, electroweak precision tests, flavor physics and dark matter. In addition in \Secref{sec:pheno} we also present predictions relevant for Higgs searches and comment briefly on the consequences for SUSY searches.
Our conclusion is given in \Secref{sec:conclusions} and in the Appendices we present, for completeness, relevant expressions for sparticle mass matrices and the effective potential used in our analysis.



\section{The scale-invariant NMSSM}
\label{sec:NMSSM}

We consider a simple extension of the MSSM, where an extra chiral singlet superfield $S$ is added to the usual MSSM field content. The superpotential couplings of the singlet are constrained by a $\mathbb{Z}_3$ symmetry which forbids linear and quadratic terms. This leads to the 
superpotential of the scale-invariant NMSSM
\begin{equation}
W_{\rm NMSSM}=\lambda S H_{u}H_{d}+\frac{\kappa}{3} S^3,\label{WNMSSM}
\end{equation}
where $\lambda$ and $\kappa$ are dimensionless couplings. 
Including contributions from soft SUSY breaking and the $D$-terms, the resulting scalar potential reads
\begin{multline}
V = (m_{H_{d}}^2+\lambda^2 S^2)|H_{d}|^2+(m_{H_{u}}^2+\lambda^2 S^2)|H_{u}|^2+\lambda^2|H_{d}H_{u}|^2+m_S^2 |S|^2+\kappa^2|S|^4\\
+ [(a_{\lambda} S+\lambda\kappa S^2)H_{u}H_{d}+\frac{a_{\kappa}}{3} S^{3}+h.c.]+V_{D},
\label{fullpotential}
\end{multline}
where $m_{H_u}, m_{H_d}, m_S, a_\lambda$ and $a_\kappa$ are soft mass parameters. Furthermore, $V_{D}$ denotes the $D$-term contribution to the potential,
\begin{equation}
V_{D}=V_{U(1)_Y}+V_{SU(2)_W}=\frac{(g_1^2+g_{2}^2)}{8}(|H_{u}|^2-|H_{d}|^2)^2+\frac{g_{2}^2}{8}|H_{d}^{\dagger}H_{u}|^2,
\end{equation}
where $g_1$ is the U(1)$_Y$ and $g_2$ is the SU(2) gauge coupling.
To ensure $CP$-conserving VEVs, we shall assume that the Lagrangian of the Higgs sector is $CP$-invariant. 
By field redefinitions, it is then possible to choose $\lambda, a_{\kappa}, v_{u}$ and 
$v_d$ real and positive, while the rest of the Higgs-sector parameters are real but can have both signs. For completeness, let us mention that the Yukawa couplings $y_t>0$ and $y_b>0$ can similarly be chosen real and positive.

The soft SUSY breaking terms are generated at a scale $\Mmess$, below which they receive loop corrections via renormalization group (RG) running.
The scale $\Mmess$ could be associated with the mass of messenger fields which communicate SUSY breaking from a hidden sector to the visible sector. Alternatively, $\Mmess$ could be a cutoff below which NMSSM fields emerge from an underlying strongly coupled theory as, for example~in the models of \cite{Gherghetta:2003he,Harnik:2003rs,Sundrum:2009gv,Gherghetta:2011wc}. For definiteness, we shall refer to $\Mmess$ as the messenger scale in the following. Our aim is to identify a region of parameter space with minimal fine-tuning for the scale-invariant NMSSM. In order to limit loop corrections to soft masses from RG running, we shall therefore focus on a relatively low messenger scale $\Mmess = 20 \TeV$ (though we also comment on the cases $\smash{\Mmess=100 \TeV}$ and $1000$ TeV). 

The NMSSM provides an additional tree-level contribution to the Higgs mass which is enhanced for large values of $\lambda$ (see below). Due to the fast running of $\lambda$, values $\lambda \gtrsim 0.7$ at the electroweak scale lead to a Landau pole below the GUT scale. In order to accommodate larger $\lambda$ and thus a sizable tree-level contribution to the Higgs mass, we will only require perturbativity of the couplings up to the messenger scale. This is in the spirit of models of $\lambda$SUSY~\cite{Barbieri:2006bg,Hall:2011aa} and allows for values $\lambda\lesssim 2.3$ at the electroweak scale. Note that some form of UV completion is required to kick in before the Landau pole in $\lambda$ (or any of the other couplings). The cutoff associated with this UV completion would have to coincide with the messenger scale $\Mmess$ if a Landau pole occurs shortly above $\Mmess$ (e.g.~for $\lambda \approx 2.3$) but otherwise this is not necessarily the case.

In order to account for quantum fluctuations we make use of effective field theory  in the form of 
effective action methods that are based on the run-and-match procedure by which all relevant parameters are run with the RG scale, for scales greater than the masses involved in the theory. When the RG scale goes through a particular mass threshold, heavy fields decouple, leaving threshold effects that are used to match the effective field theory above and below the mass threshold. The effective action for the Higgs scalars, through a loop expansion, can be written as:
\begin{equation}
\mathcal{S}_{eff}=\int d^4x \left\{\sum_{n=0}^{\infty}Z^n_i\partial_{\mu}\phi_i^{\dagger}\partial^{\mu}\phi_i-\sum_{n=0}^{\infty}V_n\right\},
\label{effS}
\end{equation}
where $\phi_i=H_u, H_d, S$,  the $Z^n_i$ account for wave-function renormalization, and the loop calculation involves a mass parameter $\mu_r$ known as the renormalization scale. 
The well-known Coleman-Weinberg~\cite{Coleman:1973jx} expression for the one-loop effective potential in the $\overline{DR}$ scheme is given by
\begin{equation}
\label{Coleman-Weinberg}
 V_{\rm 1} =\frac{1}{64\pi^2} \, {\rm STr} \, M^4\left[ \log\left(\frac{M^2}{\mu_r^2}\right)-\frac{3}{2}\right],
\end{equation}
where the trace runs over all particles in the effective theory, $M$ denotes their masses and all dimensionless and dimensionful couplings of the theory are evaluated at $\mu_r$.  The dominant contributions to $V_{\rm 1}$ can be found in the Appendix.  
A sensible choice for the renormalization scale $\mu_r$ that makes the approximation more scale-independent
and minimizes large logarithms arising from the top/stop sector is $\mu_r=\msoft\equiv\sqrt{m_{Q_3}m_{u_3}}$, where $m_{Q_3}$ and $m_{u_3}$ are the soft-breaking masses of the stop sector. Therefore, above the scale $\msoft$ we have an effective supersymmetric theory where all couplings can be determined at any scale $\msoft\lesssim Q\lesssim \Mmess$ using supersymmetric RG running.
In this way, potentially divergent logarithms that involve the messenger scale $\Mmess$ are taken into account by integrating the SUSY RG equations, while in principle all finite terms are accounted for by the effective action. Given the potentially large values of the couplings $y_t$, $\lambda$, $\kappa$ at the SUSY scale, we expect that the major contributions to the effective action come from the stop/top and Higgs-singlet sector. 

Minimizing the effective potential (including 1-loop corrections) with respect to the neutral components of $H_u$, $H_d$ and $S$,  we find that the following minimization conditions must be satisfied
\begin{eqnarray}
\widehat{m}_{H_{u}}^2+\lambda^2 v_S^2-(a_{\lambda} v_S+\lambda\kappa v_S^2)\frac{1}{\tan\beta}-\frac{m_{Z}^2}{2}\cos2\beta+\lambda^2 v^2 \cos^2\beta&=&0, \label{min1}\\
\widehat{m}_{H_{d}}^2+\lambda^2 v_S^2-(a_{\lambda} v_S+\lambda\kappa v_S^2)\tan\beta+\frac{m_{Z}^2}{2}\cos2\beta+\lambda^2 v^2 \sin^2\beta&=&0, \label{min2}\\
\widehat{m}_S^2-\lambda\kappa v^2 \sin2\beta+2\kappa^2 v_S^2 +\lambda^2 v^2-\frac{a_{\lambda}v^2}{2v_S}\sin2\beta+a_{\kappa}v_S&=&0,\label{smin} 
\end{eqnarray}
where
\begin{equation}
\widehat{m}_{H_{u}}^2 \equiv m_{H_{u}}^2 + \frac{d}{d v_u^2}  V_{\rm 1}, \qquad \widehat{m}_{H_{d}}^2 \equiv m_{H_{d}}^2 + \frac{d}{d v_d^2}  V_{\rm 1}, \qquad \widehat{m}_S^2 \equiv m_S^2 + \frac{d}{d v_S^2}  V_{\rm 1} ,
\label{modmhu2etc}
\end{equation}
and $v^2=v_{u}^2+v_d^2$, $\tan\beta=v_u/v_d$, $m_{Z}^2=\frac{1}{2}(g_1^2+g_2^2) \, v^2$ and all parameters are evaluated at the renormalization scale $\mu_r$. From the first two equations, we can derive the following relations:
\begin{eqnarray}
\lambda^2v^2&=&2\frac{(a_{\lambda}v_S+\lambda\kappa v_S^2)}{\sin2\beta}-\widehat{m}_{H_{u}}^2-\widehat{m}_{H_{d}}^2-2\lambda^2v_S^2,\label{v1}\\
m_{Z}^2&=&\frac{\widehat{m}_{H_{u}}^2-\widehat{m}_{H_{d}}^2}{\cos 2\beta}-\widehat{m}_{H_{u}}^2-\widehat{m}_{H_{d}}^2-2\lambda^2v_S^2.\label{v2}
\label{modmincon}
\end{eqnarray}
Notice that the second condition (\ref{v2}) is similar to the one obtained in the MSSM if we identify $\mu \equiv \lambda v_S$ and neglect Higgs-singlet contributions to the effective action.
As can be seen from Eq.~(\ref{effS}), the fields $H_u$, $H_d$ and $S$ do not yet have properly normalized kinetic terms. Absorbing  the wavefunction renormalization factors $Z_{H_u}$, $Z_{H_d}$ and $Z_{S}$ into the fields, the VEVs of the properly normalized fields are related to the VEVs $v_u$, $v_d$ and $v_S$ found from \eqref{min1}-\eqref{smin} according to $\smash{v_{u, {\rm norm.}} =  v_u \sqrt{Z_{H_u}}}$  and similarly for $v_d$ and $v_S$. In the following, we will drop the subscript `norm.' and $v_u$, $v_d$ and $v_S$ will instead refer to the VEVs of the properly normalized fields. We then require that $v =174 \GeV$. 

Expanding the neutral components of the Higgs fields around their VEVs,
\begin{equation}
H_u^0=v_u+\frac{h_u+i h_{u,I}}{\sqrt{2}}, \qquad H_d^0=v_d+\frac{h_d+i h_{d,I}}{\sqrt{2}}, \qquad S=v_S+\frac{s+i s_{I}}{\sqrt{2}}, 
\end{equation}
where the subscript $I$ denotes the imaginary part,
one finds the mass matrix of $CP$-even states:
\begin{equation}
(\mathcal{M}^2_{CP-even})_{ij}=\frac{(M^2_{CP-even})_{ij} +\Delta M^2_{ij}}{\sqrt{ Z_iZ_j}}~.
\label{CPevenMassMatrixLoops}
\end{equation}
Here $M^2_{CP-even}$ denotes the tree-level contribution which in the basis $(h_u,h_d,s)$ is given by the matrix
\begin{eqnarray}
\label{CPevenMassMatrixTree}
{\footnotesize \left(
\begin{array}{ccc}
m_Z^2 \sin^2\beta+m_A^2\cos^2\beta  &\left(\lambda^2 v^2-\frac{1}{2}\left(m_Z^2+m_A^2\right)\right)\sin 2\beta & 2\lambda\mu v \sin\beta-a_{\lambda} v \cos\beta-2\kappa\mu v \cos\beta   \\
.  & m_A^2 \sin^2\beta+m_Z^2\cos^2\beta &  2\lambda\mu v \cos\beta-a_{\lambda} v \sin\beta-2\kappa\mu v \sin\beta   \\
. & . & \lambda^2 v^2+ m_S^2+6 \frac{\mu^2}{\lambda^2}\kappa^2-\lambda\kappa v^2 \sin 2\beta +2 a_k \frac{\mu}{\lambda}
\end{array}
\right) . }
\end{eqnarray}
We adopt the convention that dotted entries are obtained from the fact that the matrix is symmetric and, for convenience, we have defined $\mu \equiv \lambda v_S$ and 
\begin{equation}
m_{A}^2 \equiv \frac{2}{\sin2\beta} (a_{\lambda}v_S+\lambda\kappa v_S^2)\, .
\end{equation}
Furthermore, $\Delta M^2$ denotes contributions from loop corrections to the effective potential which are given by
\begin{equation}
\Delta M^2_{ij}=\frac{1}{2}\frac{\partial^2 V_{1}}{\partial v_i \partial v_j} \qquad {\rm for}\; i\neq j\;, \qquad \Delta M^2_{ii}=\frac{1}{2}\frac{\partial^2 V_{1}}{\partial v_i^2} -\frac{\partial V_{1}}{\partial v_i^2}  \, .
\label{CPevenMassMatrix}
\end{equation}

The eigenvalues $m_i^2(\mu_r)$ 
of the mass matrix \eqref{CPevenMassMatrixLoops} are running masses evaluated at the renormalization scale $\mu_r$.  From these, the physical pole masses are obtained by solving the pole equation
\begin{equation}
{\rm det}\left[p^2\delta_{ij}-(m^2_{i}(\mu_r)+\Delta\Pi_i(p^2))\right]=0~,
\end{equation}
where $\Delta\Pi_i(p^2)\equiv \Pi_{i}(p^2)-\Pi_i(0)$ and $\Pi_i(p^2)$ are the $\overline{DR}$-renormalized self energies.

It is useful to consider a basis 
in which the electroweak VEV is associated with only one linear combination of states $h'$. To this end, we rotate
\bea
\left(
\begin{array}{ccc}
h'  \\
H' \\
s
\end{array}
\right)
&=&
\left(
\begin{array}{ccc}
\sin\beta  & \cos\beta & 0   \\
-\cos\beta  & \sin\beta & 0 \\
0              &       0    &      1  
\end{array}
\right)
\left(
\begin{array}{ccc}
h_u \\
h_d \\
s
\end{array}
\right).
\label{eq:SMHiggsBasis}
\eea
In this basis, the diagonal element of the tree-level mass matrix associated with $h'$ becomes
\begin{equation}
m_{h'h'}^2 \equiv m^2_Z\cos^2 2\beta+\lambda^2 v^2\sin^2 2\beta~.
\label{mhbound}
\end{equation}
This constitutes an upper bound on the lightest eigenvalue of the $CP$-even mass matrix at tree-level. The second term, which is not present in the MSSM, implies that the Higgs mass can be larger than the $Z$-boson mass already at tree level. In comparison to the MSSM, this alleviates the need for heavy stops to raise the Higgs mass to 126\GeV.

Note that the bound is saturated (so that Eq.~\eqref{mhbound} \emph{is} the Higgs mass) if the lightest $CP$-even state is purely $h'$. Since in this basis, the VEV is only carried by $h'$, such a Higgs would couple to SM fields just like the SM Higgs. Given the uncertainties in the production cross sections and branching fractions of the Higgs (and the hint for an excess in $h\rightarrow \gamma \gamma$), a non-vanishing admixture of $h'$ with the orthogonal state $H'$ and the singlet $s$ is still permissible and typically present. This reduces\footnote{The state identified with the Higgs might not be the lightest $CP$-even state (see, for example~\cite{Agashe:2012zq}). In this case, the admixture \emph{raises} the Higgs mass. We do not find any such points in our numerical scan and will therefore not consider this case further.} the Higgs mass relative to Eq.~\eqref{mhbound}. In addition, loop corrections may either increase or reduce the Higgs mass.



\section{Numerical Analysis}
\label{sec:NumAnal}
Before proceeding to the discussion of the imposed bounds and results in detail, we give a brief description of how the numerical scan is performed. Readers who are mainly interested in our results could skip this section. 

As motivated in the Introduction, we decouple all sparticles from the first and second generation as well as third-generation sleptons. The remaining parameter space of the scale-invariant NMSSM is characterized by the 16 parameters given in \Tabref{tab:InputParams}. Note that we use the VEVs $v_u$, $v_d$ and $v_S$ (or $v$, $\tan \beta$ and $\mu$) as input parameters instead of the soft masses $m_{H_u}^2$,  $m_{H_d}^2$ and $m_S^2$. The latter can be obtained from the former by means of the minimization conditions \eqref{min1} - \eqref{smin}. 
\begin{table}[tb]\centering
\begin{tabular}{l|c||l|c||l|c}\hline\hline
$\tan\beta$ & $\tan\beta>0.08$&
$m_{ Q_3}$&$\Delta_{\tilde g} m_{ Q_3}<m_{ Q_3}<5\TeV$&
$M_1$ & $0<M_1<8\TeV$\\
$\mu$ & $|\mu|<1\TeV$ &
$m_{ u_3}$&$\Delta_{\tilde g} m_{ u_3}<m_{ u_3}<5\TeV$&
$M_2$ & $0<M_2<8\TeV$\\
$\lambda$ & $0<\lambda<3$&
$m_{ d_3}$&$0<m_{ d_3}<8\TeV$&
$M_3$ & $0.5\TeV<M_3<8\TeV$\\
$\kappa$ & $|\kappa|<2.75$&
$A_t$ & $|\Delta_{\tilde g} A_t|<|A_t|<5\TeV$&
$v$&$ 174\GeV$\\
$A_\lambda$ & $|A_\lambda|<2\TeV$&
$A_b$ & $|A_b|<8\TeV$&
$\Mmess$&$20, 100, 1000 \TeV$\\
$A_\kappa$ & $|A_\kappa|<1\TeV$ &
&&&\\\hline\hline
\end{tabular}
\caption{\em The ranges of the input parameters (defined at the renormalization scale $\msoft$) which we  use in the numerical scan. Here $A_\lambda\equiv a_\lambda/\lambda$, $A_\kappa\equiv a_\kappa/\kappa$  and $\Delta_{\tilde g} \xi$ denotes the one-loop contribution of gluinos to the soft mass parameters $\xi=A_t, m_{Q_3}, m_{u_3}$. We choose $a_\kappa >0$ without loss of generality as discussed in \Secref{sec:NMSSM}. 
\label{tab:InputParams}}
\end{table}
After fixing the Higgs VEV at ${v=174\GeV}$, the messenger scale at $\Mmess=20\TeV$  (and all SM masses and couplings at their measured values),
we perform a numerical analysis by sampling points in the remaining 14-dimensional parameter space defined at the renormalization scale $\msoft=\sqrt{m_{Q_3} m_{u_3}}$. To this end, we employ a Markov Chain Monte Carlo using the Metropolis-Hastings algorithm with simulated annealing. We scan linearly in all parameters, i.e. we prefer large values in the scan, which helps to find the boundary of the region with small fine-tuning. 
The likelihood function in the scan is given by the product of a Gaussian for the Higgs mass (i.e.~the state that we identify with the recently discovered resonance) centered at $126\GeV$ and a Gaussian for the fine-tuning centered at $0$, i.e.~the sampler prefers regions which have a Higgs mass close to the experimentally observed value and small fine-tuning in the electroweak VEV.\footlabel{footnote:tree}{Note that, in order to speed up the scan, we use the fine-tuning calculated from the \emph{tree-level} potential in the likelihood function. After the scan, we recalculate the fine-tuning using the effective potential including one-loop corrections.}
We will discuss the exact definition of the fine-tuning measure in \Secref{naturalness}. Let us mention that the Markov chains are relatively short and therefore there is no statistical interpretation of the scatter plots\footnote{The plots have been produced using {\em matplotlib}~\cite{Hunter:2007}.} presented in the next sections.

In addition to the hard cuts on the input parameter input given in \Tabref{tab:InputParams}, we restrict the running of the squark soft parameters by imposing\footnote{In particular, via this condition the gluino limits the size of the stop mass parameters: The gluino ``sucks'' the stop mass up as has been pointed out in \cite{Arvanitaki:2012ps}. We find values of $m_{Q_3}$ and $m_{u_3}$ up to a factor 4 smaller than the gluino mass.
This is somewhat less stringent than the ratio $m_{\tilde{g}} / m_{\tilde{t}} \lesssim 2$  given in \cite{Arvanitaki:2012ps},
because we impose the condition \eqref{eq:noTuningInSquarkParameters} and include loop corrections to the stop mass parameters from $m_{H_u}^2$ and $A_t$.
After imposing the LHC constraints on the gluino mass (see \Secref{sec:colliderlimits}), we find values of $m_{Q_3}$ and $m_{u_3}$ down to about $300 \GeV$ (for sufficiently high neutralino mass). Such small stop mass parameters still allow for small fine-tuning.\label{gluinosucks}}
\begin{equation}
\label{eq:noTuningInSquarkParameters}
| \xi(\Mmess)-\xi(\msoft) | < | \xi(\Mmess) |
\end{equation}
for $\xi=m_{Q_3}^2, m_{u_3}^2, m_{d_3}^2, A_t, A_b$.
This avoids situations where the squark soft parameters at the low scale $\msoft$
are much smaller than, for example, the loop corrections from gluinos to these parameters which would require an additional tuning. 
We do not restrict the Higgs soft parameters because they are already covered by the fine-tuning measure.

We use the program {\it NMHDECAY}~\cite{Ellwanger:2004xm,Ellwanger:2005dv,Belanger:2005kh} contained in the package {\it NMSSMTools} 3.2.1~\cite{Djouadi:2008uw} to generate the sparticle spectrum. The program is modified, however, to incorporate a low messenger scale $\Mmess$.  

In particular, {\it NMHDECAY} calculates the soft masses $m_{H_u}^2$, $m_{H_d}^2$ and $m_S^2$ from the input parameters using the minimization conditions. Loop corrections from the Higgs-singlet sector (including superpartners) to the minimization conditions are neglected in this calculation.
For large values $\lambda,\kappa\gtrsim\mathcal{O}(1)$, however,  these corrections can become sizable. In order to ensure that the results obtained from {\it NMHDECAY} are reliable, we restrict ourselves to regions in parameter space where the resulting corrections to $m_{H_u}^2$, $m_{H_d}^2$ and $m_S^2$ are relatively small. We therefore require that (cf.~Eq.~\eqref{modmhu2etc})
\begin{equation}
\left|\frac{d}{d v_u^2}  V_{1,S}\right| < \left|m_{H_u}^2\right|,\qquad
\left|\frac{d}{d v_d^2}  V_{1,S}\right|<\left|m_{H_d}^2\right|,\quad\mathrm{and}\qquad
\left|\frac{d}{d v_S^2}  V_{1,S}\right|< \left|m_S^2\right|\;.
\label{HiggsCorrLimits}
\end{equation}
The soft masses enter into our calculation of the fine-tuning measure, leading to an uncertainty of order 1 from the above condition. We believe that this is acceptable, given the uncertainties in the proper definition of the fine-tuning measure itself. In addition, 
 {\it NMHDECAY} uses its results for $m_{H_u}^2$, $m_{H_d}^2$ and $m_S^2$  in the check whether the minimum specified by the input parameters $v$, $\tan \beta$ and $\mu$ is the global minimum. We do not expect, however, that this check is significantly affected by the $\mathcal{O}(1)$-uncertainties in these soft masses. Nevertheless, a future implementation in {\it NMHDECAY}  of loop corrections  from the Higgs-singlet sector to the minimization conditions may be worthwhile. Similarly, {\it NMHDECAY} does not check whether there is a deeper minimum with VEVs $v\neq 174$ GeV and $v_S= 0$. This can be important as has recently been pointed out in Ref.~\cite{Agashe:2012zq}. 

We apply all constraints which are implemented in {\it NMSSMTools} 3.2.1, in particular several LEP and Tevatron exclusion limits. The program also calculates various flavor observables on which we impose the latest constraints. Note though that we do not try to explain the anomalous magnetic moment of the muon.  
The relic abundance of the LSP and its spin-independent proton and neutron cross sections are calculated using the program {\it MicrOMEGAs} 2.4.5~\cite{Belanger:2005kh,Belanger:2010gh}. We require that the universe is not overclosed and that the latest XENON100 bounds~\cite{Aprile:2012nq} on the cross sections for a given relic abundance are satisfied. 
We take into account the constraints on oblique corrections and require that the Peskin-Takeuchi parameters, $S$ and $T$~\cite{Peskin:1991sw} are within their $2\sigma$ limits. 
In addition, {\it NMHDECAY} provides the Higgs couplings normalized to the SM as well as the normalized cross sections of the most important LHC Higgs detection channels. 
We impose the experimental bounds on the different Higgs cross sections. 
Furthermore, we restrict the scan to small fine-tuning by imposing $\Sigma^v<200$ on the fine-tuning measure calculated from the tree-level\footref{footnote:tree}
minimization conditions. For the final results, we recalculate the fine-tuning measure including the dominant stop-top corrections to the effective potential. Note that we neglect the corrections from the Higgs-singlet sector, on the other hand, as they are restricted to be small according to \eqref{HiggsCorrLimits}.
In order to generate data points for $\Mmess=100 \TeV$ and $1000 \TeV$, we use the generated points for $\Mmess=20\TeV$, 
and recalculate the constraints and fine-tuning with the new $\Mmess$. Note that some points are excluded for larger $\Mmess$ because they have a Landau pole between $20\TeV$ and the bigger $\Mmess$.



\section{Naturalness Constraints}
\label{naturalness}

\subsection{Electroweak scale tuning: Large $\lambda$ helps}

We shall now determine to what extent the (scale-invariant) NMSSM maintains naturalness in light of the LHC results which place 
stringent lower limits on the sparticle masses. As can be seen from Eqs. (\ref{v1}) and (\ref{v2}), soft masses of order the TeV scale require a delicate cancellation in order to obtain the correct electroweak scale $v=174$ GeV and thereby necessitate some amount of fine-tuning.

In order to quantify this fine-tuning, we use the measure proposed in Ref.~\cite{Barbieri:1987fn}, which determines how small deviations in the input parameters of the theory affect the electroweak scale:
\begin{equation}
\Sigma^v \equiv \max_i \left| \frac{d \log v^2}{d \log \xi_i(\Mmess) }\right|  \, .
\label{sigmavev}
\end{equation}
Here the input parameters $\xi_i=(m^2_{H_u}, m^2_{H_d}, m^2_S, \lambda, \kappa, a_{\lambda}, a_{\kappa},  m^2_{Q_3}, m^2_{u_3}, m^2_{d_3}, A_t, A_b, M_1, M_2, M_3)$ are 
defined at the messenger scale $\Mmess$. This set consists of the input parameters on which the Higgs VEV $v$ has a significant dependence.
For the plots, we shall typically require that $\Sigma^v < 20$, corresponding to a fine-tuning better than  5\%.

Using the minimization conditions (\ref{min1}--\ref{smin}), we can solve for the VEVs $v_u, v_d$ and $v_S$. We  take the one-loop corrections to the effective potential due to third-generation quarks/squarks into account and choose $\msoft= \sqrt{m_{Q_3} m_{u_3}}$ as the renormalization scale (cf. the discussion in~\Secref{sec:NMSSM}). The VEVs $v_u,v_d$ and $v_S$ are then functions of running masses and couplings evaluated at this scale. We shall collectively denote these parameters by $\xi_i(\msoft)$. In order to evaluate the fine-tuning measure \eqref{sigmavev}, we differentiate the minimization conditions (\ref{min1}--\ref{smin}) with respect to the $\xi_i(\msoft)$ and solve the linearly-coupled system of equations for the derivatives $dv^2/d\xi_i(\msoft)$. We solve the RG equations in the leading-log approximation to obtain an analytical expression for the $\xi_i(\msoft)$ in terms of the parameters $\xi_i(\Mmess)$ defined at the messenger scale. To this end, we take all relevant contributions at one-loop order into account  and also include the two-loop corrections from gluinos that feed into $m^2_{H_u}$,  $m^2_{H_d}$ and $a_{\lambda}$. Using the chain rule, we can then calculate the fine-tuning measure according to:
\begin{equation}
\Sigma^v = \max_i \left| \sum_j \frac{\xi_i(\Mmess)}{v^2} \frac{d v^2}{d\xi_j(\msoft)} \frac{d\xi_j(\msoft)}{d\xi_i(\Mmess)} \right| \, .
\label{sigmavev2}
\end{equation}

Let us briefly mention a subtlety when solving the RG equations: Since $\lambda$ and $\kappa$ are typically large and accordingly run quickly, we do not use a leading-log approximation for these couplings. Instead, we solve their RG equations numerically and integrate them properly in the RG equations of the other couplings. 
Furthermore, as $\lambda$ and $\kappa$ are renormalized multiplicatively, we find that
\begin{align}
\frac{d\ln \lambda(\msoft)}{d\ln\lambda(\Mmess)}&\approx 1\;,&
\frac{d\ln \kappa(\msoft)}{d\ln\kappa(\Mmess)}&\approx 1\;,&
\frac{d\ln \lambda(\msoft)}{d\ln\kappa(\Mmess)}&\approx 0\;,&
\frac{d\ln \kappa(\msoft)}{d\ln\lambda(\Mmess)}&\approx 0\;.
\end{align}
Using these relations and the fact that the dependence of the electroweak scale on dimensionful parameters is already taken into account by other terms in the fine-tuning measure (\ref{sigmavev}), we will approximate the logarithmic derivative with respect to $\lambda(\Mmess)$ by that with respect to $\lambda(\msoft)$ and similarly for $\kappa$.

It is instructive to consider regions of parameter space where large stop mass parameters $m_{Q_3}^2$, $m_{u_3}^2$ and $A_t$ 
at the messenger scale dominate the fine-tuning measure. In particular, the stops affect the Higgs mass parameter $m_{H_u}^2(m_{soft})$ via the RG evolution. Using the leading-log approximation, we find
\begin{equation}
m^2_{H_u}(\msoft) =  m^2_{H_u}(\Mmess)  - \frac{3 y_t^2}{8 \pi^2}\left [ m^2_{Q_3}(\Mmess) +m^2_{u_3}(\Mmess) +A_{t}^2(\Mmess)  \right] \log\left[\frac{\Mmess}{\msoft}\right] +\dots
\label{mhutuning}
\end{equation}
If the contribution from the stop sector is significant,  the value of $ m^2_{H_u}(\Mmess)$ has thus to be tuned in order to obtain a value of $m^2_{H_u}(\msoft)$ not much larger than the electroweak scale.  
Note that in our scan of parameter space for $\Mmess=20$ TeV,  we find that $m^2_{H_u}$ dominates the VEV tuning for approximately half of the points,  while the next relevant parameters that dominate the tuning are $\lambda$ and $a_{\lambda}$. When the messenger scale is increased to $\Mmess=100$ TeV, $m^2_{H_u}$ still dominates the VEV tuning for approximately half of the points, but the next dominating contributions\footnote{This might originate from our way of generating the data points for $\Mmess=100(1000)\TeV$ compared to $\Mmess=20\TeV$.} are now $M_2$ and $M_3$.

We can obtain an upper bound on the stop mass parameters from naturalness by looking at how they affect the fine-tuning measure. Neglecting correction from the Coleman-Weinberg potential, we in particular find that
\begin{equation}
\left| \frac{d \log v^2}{d \log \xi_t(\Mmess)}\right| \, \approx \,  \left| \frac{3 y_t^2}{8 \pi^2}\frac{\xi_t(\Mmess)}{v^2}\log\left[\frac{\Mmess}{\msoft} \right] \times \frac{dv^2}{dm^2_{H_u}(\msoft)} \right| \, \lesssim \, \Sigma^v\label{stoptuning}
\end{equation}
for $\xi_t =(m^2_{Q_3},m^2_{u_3},  A^2_t)$. This means that, for a given amount of fine-tuning, the parameters $\xi_t$ and thus the stop masses have an upper limit which depends on the size of the derivative $dv^2 / dm^2_{H_u}$. Note, however, that the logarithmic derivative with respect to $m^2_{H_u}(\Mmess)$ often gives a  more stringent bound on stop masses from naturalness since it effectively contains the sum of the three individual measures in Eq.~(\ref{stoptuning}).

The stops affect the electroweak scale $v$ also via finite corrections which are taken into account by the Coleman-Weinberg potential. These have been neglected in the derivation of Eq.~\eqref{stoptuning}.
From Eq.~(\ref{mhutuning}), it follows that as the ratio $\Mmess/\msoft$ increases, the contributions from the RG running become more important compared to these finite corrections. For our particular choice of parameters, we estimate that for $\Mmess\gtrsim 100$ TeV, the latter can be neglected in the calculation of the fine-tuning measure. 

\begin{figure}[t]
\begin{subfigure}{0.49\linewidth}
\includegraphics[width=\linewidth]{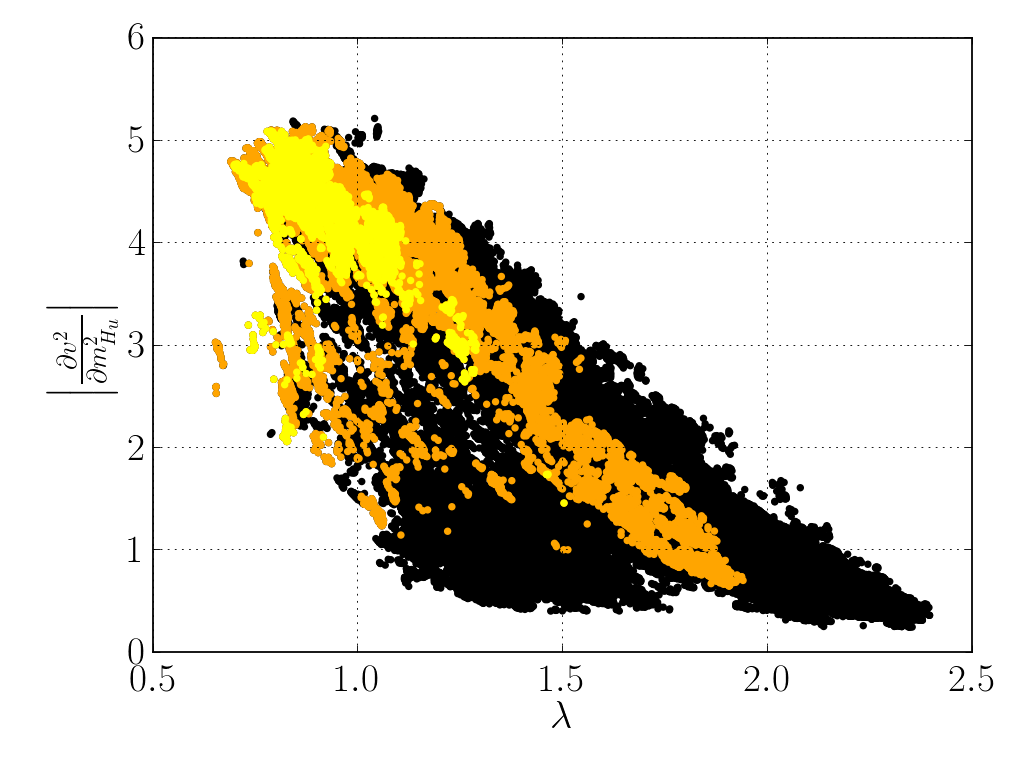}
\caption{\label{fig:dv2dmhu2vslambda}}
\end{subfigure}
\hfill
\begin{subfigure}{0.49\linewidth}
\includegraphics[width=\linewidth]{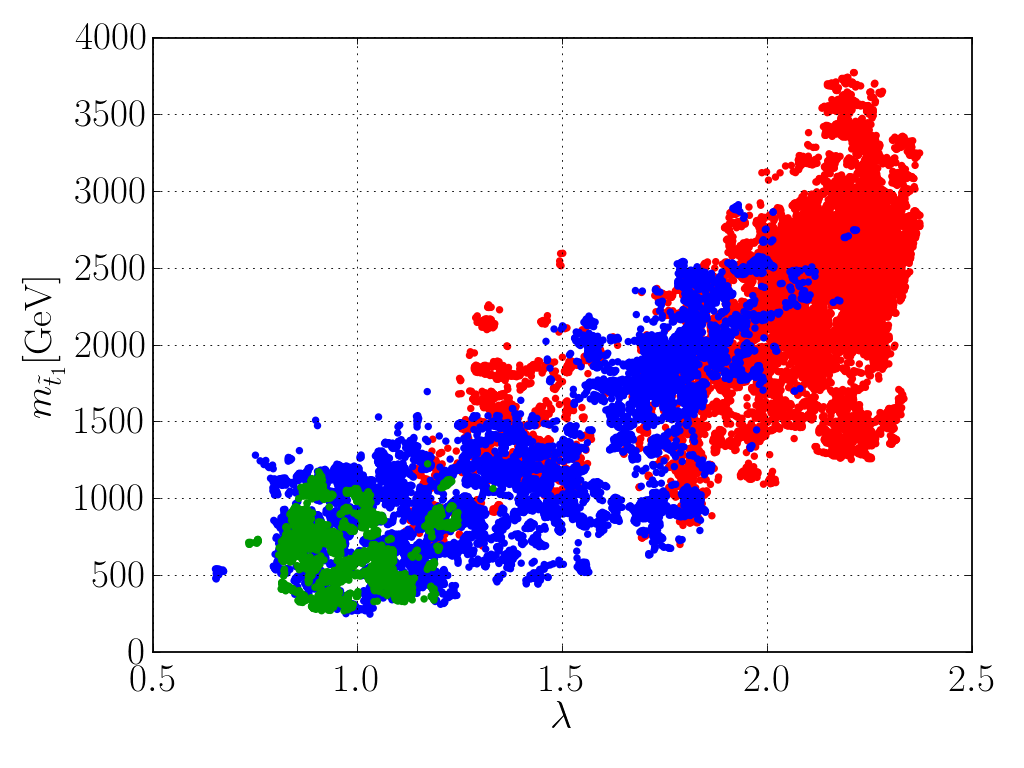}
\caption{\label{fig:mstop1vslambda}}
\end{subfigure}
\caption{\em Scatter plots of (a) the tree-level derivative $|dv^2 / dm^2_{H_u}|$, and (b) the lightest stop mass $m_{\tilde{t}_1}$, as a function of the Higgs-singlet coupling $\lambda$. In (a) the black, orange, yellow points correspond to $\Mmess=20,100, 1000\TeV$, respectively. All points in (b) have $\Mmess=20\TeV$ and a tuning in the Higgs VEV better than $5\%$. In (b) the green points have a combined tuning (cf.~\Secref{sec:HiggsMassFineTuning})
better than $5\%$, i.e. $\Sigma^h\Sigma^v<20$, for the blue points it is between $1\%$ and $5\%$, while for the red points it is worse than $1\%$.
The derivative $dv^2 /dm^2_{H_u}$ is suppressed for larger values of $\lambda$, allowing for $m_{\tilde{t}_1}$ to become as large as $2.5 \TeV$ for a combined tuning better than 1$\%$. All points satisfy the constraints discussed in \Secref{sec:collider}.}
\label{fig:dv2dmhu2}
\end{figure}

Let us now discuss the derivative $dv^2 / dm^2_{H_u}$ in more detail.  We first consider the situation in the MSSM.  Neglecting loop corrections to the effective potential,  we find in the limit $\tan\beta\gg 1$ that
\begin{equation}
\frac{dv^2}{dm^2_{H_u}(\msoft)}=-2\frac{v^2}{m_Z^2}+\mathcal{O}\left(\frac{1}{\tan\beta}\right).
\end{equation}
Thus there is no freedom in the MSSM to suppress this derivative and thereby reduce the tuning once the stop soft terms become large. In the NMSSM, on the other hand, we find in the limit $\lambda\gg 1$ that (again using the tree-level potential) 
\begin{equation}
\frac{dv^2}{dm^2_{H_u}(\msoft)}= \frac{\kappa}{\lambda^3} \cot 2\beta+\mathcal{O}\left(\frac{1}{\lambda^4}\right)\,.
\label{dv2dmhu2NMSSM}
\end{equation}
The derivative is thus suppressed for large $\lambda$.  This effect can be observed in \Figref{fig:dv2dmhu2vslambda}, where we present a scatter plot of the derivative as a function of $\lambda$ (calculated using the tree-level potential). The black points have a messenger scale $\Mmess=20\TeV$, while the orange (yellow) points have $\Mmess=100 \, (1000)\TeV$. There are fewer points for larger $\Mmess$ because some points are excluded by a Landau pole below $\Mmess$. 
The suppression at large $\lambda$ allows for larger stop masses compared to the MSSM for a given amount of fine-tuning. In Fig.~\ref{fig:mstop1vslambda}, we show a scatter plot of the stop mass $m_{\tilde{t}_{1}}$ versus $\lambda$, where all points have a fine-tuning better than 5$\%$. Indeed, the maximal value of $m_{\tilde{t}_{1}}$ that we find grows with $\lambda$, corresponding to the suppression of the derivative $dv^2 / dm^2_{H_u}$ with $\lambda$.
In particular, 
naturalness can be maintained for stop masses $m_{\tilde{t}_{1}}\gtrsim  1 \TeV$ as opposed to the MSSM where lighter stops $m_{\tilde{t}_1}\lesssim 600 \GeV$ are required.  Note that when $\kappa\sim \lambda$, the suppression in \eqref{dv2dmhu2NMSSM} becomes $1/\lambda^2$, and agrees with the result in \cite{Perelstein:2012qg}.

In addition, we find that the stop/top-contributions to the Coleman-Weinberg potential alleviate the fine-tuning. This has been noticed before in the context of the MSSM \cite{Cassel:2010px}. Indeed, these corrections increase the quartic Higgs coupling, resulting in a reduced sensitivity of the electroweak scale in analogy to the case of large $\lambda$. We find that on average, this effect reduces the fine-tuning measure by $10\%-20\%$ compared to the case when only the tree-level potential is taken into account. 

This observation, however, implies a caveat: We have neglected the contribution from the Higgs-singlet sector to the Coleman-Weinberg potential in the calculation of the VEV fine-tuning measure. In the region of large $\lambda$, this contribution is potentially important and could modify the suppression of the derivative $dv^2 / dm^2_{H_u}$ with $\lambda$.
This is indeed what we expect in the decoupling (or SM) limit when the Higgs potential can be described by a one-Higgs-doublet model: ${V \sim - m_{h,{\rm eff}}^2 h^2 + \lambda_{\rm eff} h^4}$. Here $m_{h,{\rm eff}}^2$ is a combination of the dimensionful parameters in the potential, whereas $\lambda_{\rm eff}$ is determined by the gauge couplings and $\lambda$. Both $m_{h,{\rm eff}}^2$ and $\lambda_{\rm eff}$ also receive important loop corrections. Minimizing this potential, we find that the Higgs VEV is given by
\begin{equation}
v^2 \, \sim \, \frac{m_{h,{\rm eff}}^2}{\lambda_{\rm eff}} \, .
\label{SMHiggsPotential}
\end{equation}
From this relation, we again see that large $\lambda$ can alleviate the fine-tuning problem. Indeed, increasing $\lambda$ typically also gives a larger effective quartic coupling $\lambda_{\rm eff}$. This in turn means that the dimensionful parameter $m_{h,{\rm eff}}^2$ is larger and can accordingly also tolerate larger loop corrections without affecting the Higgs VEV too much. In the decoupling limit, however, the dimensionful parameter in the potential is essentially the Higgs mass (up to an $\mathcal{O}(1)$-factor). With a fixed Higgs mass (and thus $m_{h,{\rm eff}} \sim 126 \GeV$), a larger coupling $\lambda$ does therefore not allow a larger dimensionful parameter $m_{h,{\rm eff}}$ in the potential
and should accordingly not alleviate the fine-tuning in the decoupling limit.

We believe that loop corrections from the Higgs-singlet sector may play an important role in this context. Indeed, once the VEV and mass in \Eqref{SMHiggsPotential} are fixed, the effective quartic coupling $\lambda_{\rm eff}$ is fixed as well. This means that, at large $\lambda$, an accidental cancellation between the $\lambda$-contribution to $\lambda_{\rm eff}$ and the loop corrections has to occur in order to bring $\lambda_{\rm eff}$ down to the required value. We will discuss this tuning (which can be phrased as a tuning in the Higgs mass) in some detail in the next section. Given that corrections from the (s)top sector raise the quartic coupling (or, equivalently, raise the Higgs mass), these corrections can not be responsible for this cancellation. The contribution from the Higgs-singlet sector, on the other hand, can lower the quartic coupling (or, equivalently, lower the Higgs mass)~\cite{Ellwanger:2009dp}. We therefore expect that these corrections may counteract the suppression of the derivative $dv^2 / dm^2_{H_u}$ at large $\lambda$ in the decoupling limit. Let us emphasize, however, that most of our points clearly deviate from this limit, where the potential and thus the effect of large $\lambda$ is more complicated. In addition, the presence of additional (s)particles with masses $\mathcal{O}(v)$ can lead to important non-trivial VEV-dependent contributions from the Coleman-Weinberg potential to \Eqref{SMHiggsPotential} even when the Higgs couples very SM-like. In any case, it may be worthwhile in the future to include the contributions from the Higgs-singlet sector to the Coleman-Weinberg potential in the calculation of the fine-tuning measure. 

\subsection{Higgs mass tuning: Large $\lambda$ hurts}
\label{sec:HiggsMassFineTuning}

In the preceding section, we have discussed how a large Higgs-singlet coupling $\lambda$ relieves the tuning which is required in order to obtain a Higgs VEV at the electroweak scale. 
As we will now see, however, large $\lambda$ introduces a new source of tuning associated with obtaining a Higgs mass around $126 \GeV$.

Let us first consider the limit in which the Higgs (i.e.~the state that we identify with the newly discovered resonance at $126 \GeV$) couples exactly like the SM Higgs. As discussed in \Secref{sec:NMSSM}, this is the case if the Higgs corresponds to the linear combination $h'$ in the basis \eqref{eq:SMHiggsBasis}. Its mass at tree-level is then given by \eqref{mhbound}. There are typically important loop corrections to this tree-level mass which come dominantly from the top-stop and Higgs-singlet (including superpartners) sector. We shall denote these contributions respectively by $\delta m_{h, \rm stop}^2$ and $\delta m_{h, \rm S}^2$. Neglecting subdominant loop corrections, the Higgs mass can be written as
\begin{equation}
m_h^2 \simeq m^2_Z\cos^2 2\beta+\lambda^2 v^2\sin^2 2\beta + \delta m_{h, \rm stop}^2 + \delta m_{h, \rm S}^2 \, .
\label{SMHiggsmass}
\end{equation}
As we will discuss in \Secref{sec:EWPT}, electroweak precision tests require small $\tan \beta$ if $\lambda$ is relatively large. The tree-level mass thus goes like $m_{h, \rm tree} \sim \lambda  \, v$ in this regime. In particular, it becomes \emph{too large} for $\lambda \gtrsim1$. Large loop corrections are then required to bring the Higgs mass down to $126 \GeV$. As is well known, the stop-top sector \emph{raises} the Higgs mass and is thus not suitable for this purpose. The Higgs-singlet sector, on the other hand, can give a sizable negative contribution~\cite{Ellwanger:2009dp}.

In our scan, we find values of $\lambda$ up to $2.4$ and $\tan \beta$ close to 1. The tree-level mass-squared for such a large $\lambda$ and small $\tan \beta$ becomes $m_{h, \rm tree}^2 \approx 10 \times (126 \GeV)^2$, i.e.~a factor 10 larger than what is required. In order to obtain the correct Higgs mass, a cancellation at the $10\%$-level has to occur between this part and the loop corrections. The input parameters of the theory have to be tuned accordingly. Note that this type of tuning is different from (and thus comes in addition to) the tuning which is required to obtain the correct Higgs VEV. Indeed, the tree-level mass in \Eqref{SMHiggsmass} is expressed in terms of the Higgs VEV (and $\tan \beta$). 
The tuning in the Higgs VEV only ensures that $v = 174 \GeV$ in \Eqref{SMHiggsmass}. 
The hierarchy for large $\lambda$ between  $m_{h, \rm tree}$ with this $v$ and the measured value of $126 \GeV$ necessitates an additional tuning in the Higgs mass. 

The discussion so far has focused on a Higgs that couples exactly as in the SM. The range of Higgs signal strengths which are consistent with the measurements at ATLAS and CMS, however,  still allow for some admixture of the orthogonal linear combination $H'$ and the singlet $s$ to the Higgs. In our scan, we find  that the state which we identify with the resonance at $126 \GeV$ always coincides with the \emph{lightest} $CP$-even Higgs. The admixture of $H'$ and $s$ then gives an additional \emph{negative} contribution to eq.~\eqref{SMHiggsmass} due to the usual effect of level repulsion. If the admixture of the orthogonal doublet is negligible, the contribution to the Higgs mass due to the singlet can be approximated as \cite{Larsen:2012rq, Agashe:2012zq}
\begin{equation}
\delta m_{h, \rm mix}^2 \simeq - \lambda^2 v^2  \,  \frac{\left(\frac{\lambda}{\kappa} - \sin 2  \beta  \, \left[1 + \frac{a_\lambda}{2 \lambda \kappa v_S} \right] \right)^2}{1 + \frac{a_\kappa}{4 \kappa^2 v_S} + \frac{a_\lambda \sin 2 \beta  \, v^2}{8 \kappa^2 v_S^3}} \, .
\label{mixingcorrection}
\end{equation}
This expression becomes exact in the limit $m_{h' h'}^2 / M_{33}^2 \rightarrow 0$, where $M_{33}^2$ denotes the $(3,3)$-element of the $CP$-even mass matrix given in \eqref{CPevenMassMatrixTree} and $m_{h' h'}$ is the tree-level mass in the SM-limit given in \eqref{mhbound}. We find that the approximation $m_h^2 = m_{h' h'}^2 + \delta m_{h, \rm mix}^2$ with the above mixing term reproduces the Higgs mass at tree-level (obtained from diagonalizing the $CP$-even mass matrix in \eqref{CPevenMassMatrixTree}) to better than $10\%$ for 
the vast majority of points in our scan. 
This corresponds to the fact that the admixture with the orthogonal doublet and thus its contribution to the Higgs mass is typically small in our sample of points. 

The `pull-down' of the Higgs mass due to the admixture with the orthogonal doublet and the singlet was considered in \cite{Agashe:2012zq} as a means to obtain the correct value of $126 \GeV$ in the regime of large $\lambda$. We find, however,  that this admixture modifies the mass by at most $40\%$ relative to \eqref{mhbound} for 
almost all our points.
The `pull-down' at tree-level is thus typically not a large effect. Instead, loop corrections tend to dominate and we find that $\smash{\delta m_{h, \rm stop}^2 + \delta m_{h, \rm S}^2 > 2 \times \delta m_{h, \rm mix}^2}$  for 
almost all our points. We believe that this is due to the following reasons: Firstly, in the regime of large $\lambda$, loop corrections from the Higgs-singlet sector are important (the latter were neglected in \cite{Agashe:2012zq}). Secondly, the Higgs signal strengths which were measured at ATLAS and CMS already put a significant limit on deviations from a SM Higgs. This restricts the admixture of $H'$ and $s$ to the Higgs and thus the size of $ \delta m_{h, \rm mix}^2$. 

We thus find that, at least for large $\lambda$, some accidental cancellation among different contributions to the Higgs mass is necessary in order to obtain the correct mass. It is useful to quantify this tuning in the Higgs mass in analogy to the fine-tuning measure for the Higgs VEV. We therefore define
\begin{equation}
    \Sigma^h\equiv \max_{\xi_i}\left|\frac{d \log m_h^2}{d \log\xi_i}\right| \, ,
    \label{quartictuning}
\end{equation}
where $\xi_i=(\lambda, \kappa, a_{\lambda}, a_{\kappa},  m^2_{Q_3}, m^2_{u_3}, m^2_{d_3}, A_t, A_b, M_1, M_2, M_3)$ are input parameters that affect the Higgs mass. Note that we do not include the soft masses $m_{H_{u}}^2$, $m_{H_{d}}^2$ and $m_S^2$ among these parameters. Instead, we consider a basis in which these soft masses have been replaced by the Higgs VEVs $v_u$, $v_d$ and $v_S$. We then keep the latter fixed when evaluating \eqref{quartictuning}. This allows us to separate the fine-tuning in the Higgs mass from that in the Higgs VEV. As before, a measure $\Sigma^h=20$ corresponds to a tuning of $5\%$.

\begin{figure}[t]
\begin{subfigure}{0.49\linewidth}
\includegraphics[width=\linewidth]{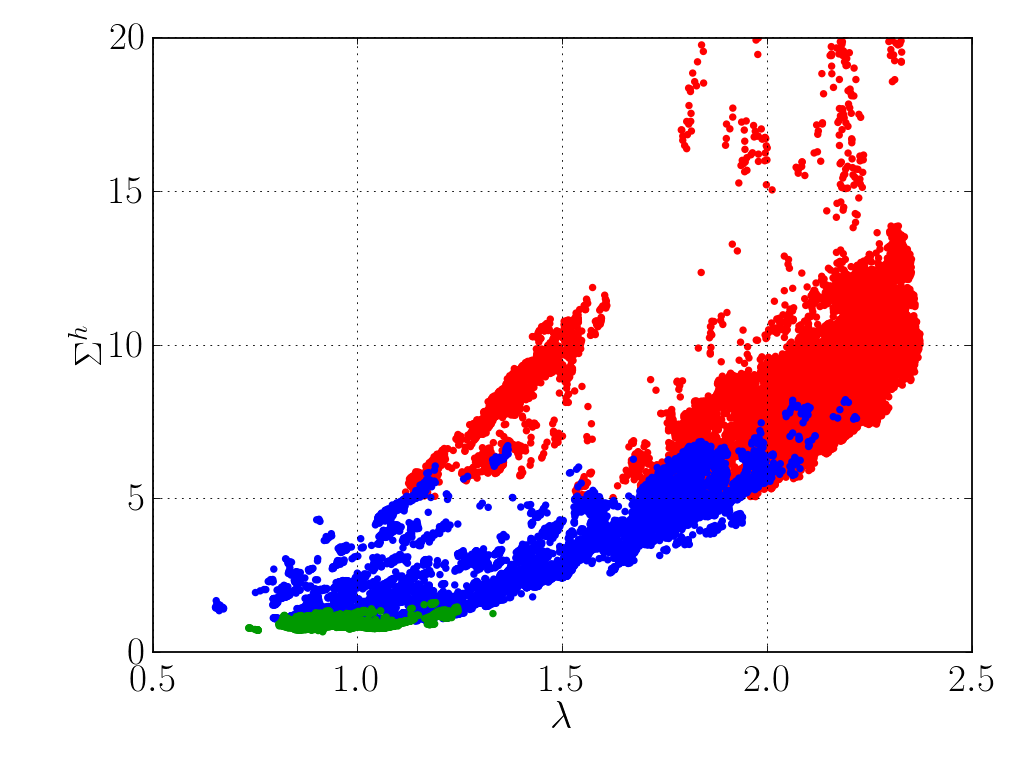}
\caption{\label{fig:higgstuning}}
\end{subfigure}
\hfill
\begin{subfigure}{0.49\linewidth}
\includegraphics[width=\linewidth]{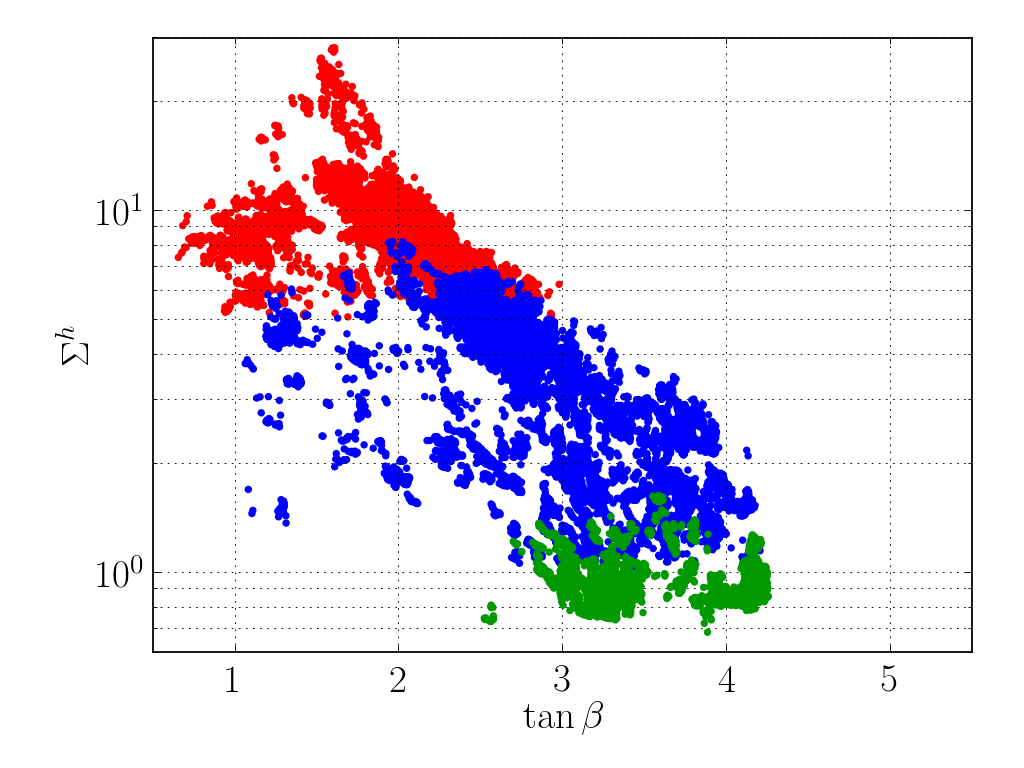}
\caption{\label{fig:higgstuningtanbeta}}
\end{subfigure}
\caption{\em The Higgs mass tuning measure $\Sigma^h$ as a function of (a) $\lambda$ and (b) $\tan \beta$ where $\Mmess=20\TeV$ and the Higgs VEV tuning is better than $5\%$. 
The green, blue and red points correspond to a combined tuning better than $5\%$, between $1\%$ and $5\%$, 
and worse than $1\%$, respectively.
Notice that the amount of tuning grows approximately like $\lambda^2$ in (a) and that it is reduced for larger $\tan \beta$ in (b). All points satisfy the constraints discussed in \Secref{sec:collider}.
}
\label{fig:totaltuninglambdakappa}
\end{figure}

We present a scatter plot of $\Sigma^h$ as a function of $\lambda$ in \Figref{fig:higgstuning}.\footnote{In order to evaluate Eq.~\eqref{quartictuning}, we calculate the tree-level contribution $m_{h,\rm tree}^2$ to the Higgs mass from the $CP$-even mass matrix in \eqref{CPevenMassMatrixTree} and the loop correction $\delta m_{h, \rm loop}^2$ using {\it NMSSMTools}. We then make the approximation $$\frac{d \log m_h^2}{d \log\xi_i  } \approx \frac{\xi_i}{m_h^2}\frac{d m_{h, \rm tree}^2}{d \xi_i}\, ,$$i.e.~we neglect the derivative of the loop correction $\delta m_{h, \rm loop}^2$ with respect to the parameters $\xi_i$ and thus a possible tuning among those $\xi_i$ which enter into $\delta m_{h, \rm loop}^2$. This approach instead only takes into account a tuning among parameters within $m_{h, \rm tree}^2$ and an accidental cancellation between $m_{h, \rm tree}^2$ and $\delta m_{h, \rm loop}^2$. Given that the tuning in the Higgs mass is typically of the latter type as discussed around Eq.~\eqref{SMHiggsmass}, we believe this to be a reasonable approximation. } Notice that the smallest tuning in the Higgs mass that we find for a given $\lambda$ grows approximately like $\lambda^2$. This can be understood from the fact that the tree-level part of the approximate mass formula \eqref{SMHiggsmass} grows like $\lambda^2$. A correspondingly larger cancellation among this tree-level part and the loop correction is therefore required at larger $\lambda$ in order to bring the Higgs mass down to $126 \GeV$. We find that less tuning is needed, on the other hand, for larger $\tan \beta$. This can be seen in \Figref{fig:higgstuningtanbeta} and is due to two effects: The tree-level mass becomes smaller for  larger $\tan \beta$ if $\lambda$ is kept fixed (assuming that $\lambda$ is so large that the term $m_Z^2 \cos^2 2 \beta$ in \eqref{SMHiggsmass} is subdominant). In addition, larger values of $\tan \beta$ require smaller values of $\lambda$ in order to satisfy constraints on the $T$-parameter (see the discussion in \Secref{sec:EWPT}). This gives another suppression on the tree-level part of the Higgs mass. 

We thus find that the Higgs mass often is a significant additional source of fine-tuning in the scale-invariant NMSSM. This type of tuning should accordingly be taken into account when assessing the naturalness of the model. In the following, we will adopt a simple possibility and multiply the tuning measures for the Higgs VEV and the Higgs mass to quantify the combined tuning.\footnote{In a statistical sense, if two quantities are not correlated, the probability involving both is $P(A\cap B)=P(A)\ast P(B)$.} This approach is justified since $\Sigma^h$ is constructed in a way that makes it independent of the tuning in the Higgs VEV as discussed before.

\begin{figure}[t]
\begin{subfigure}{0.49\linewidth}
\includegraphics[width=\linewidth]{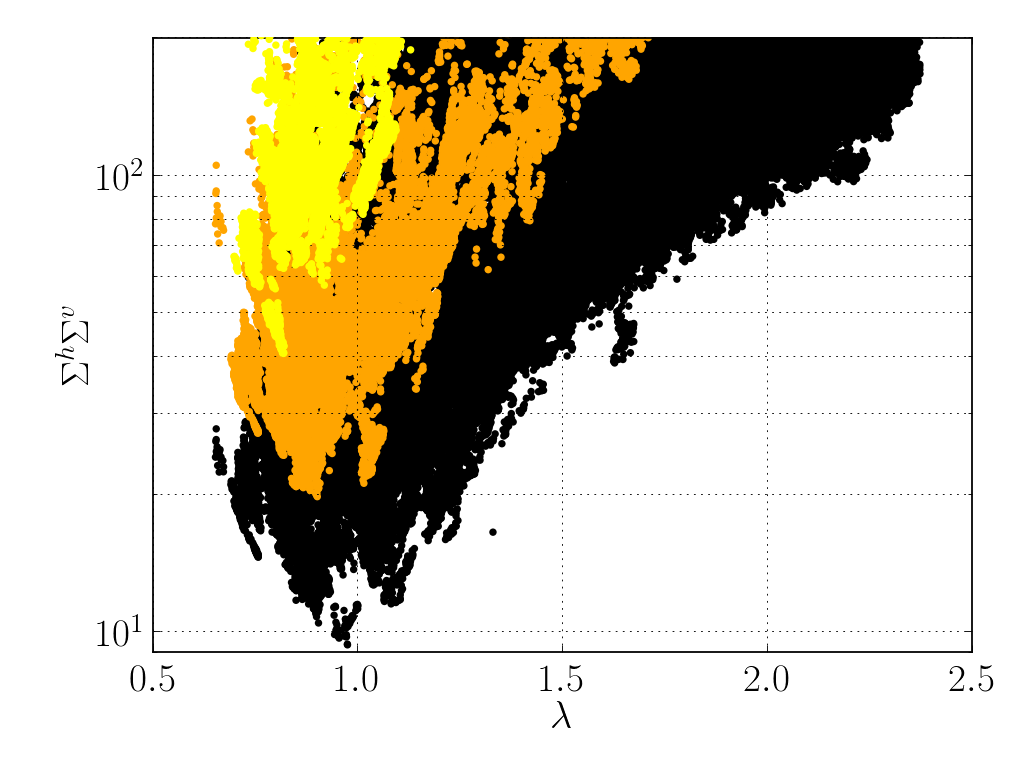}
\caption{\label{fig:totFTvslambda}}
\end{subfigure}
\hfill
\begin{subfigure}{0.49\linewidth}
\includegraphics[width=\linewidth]{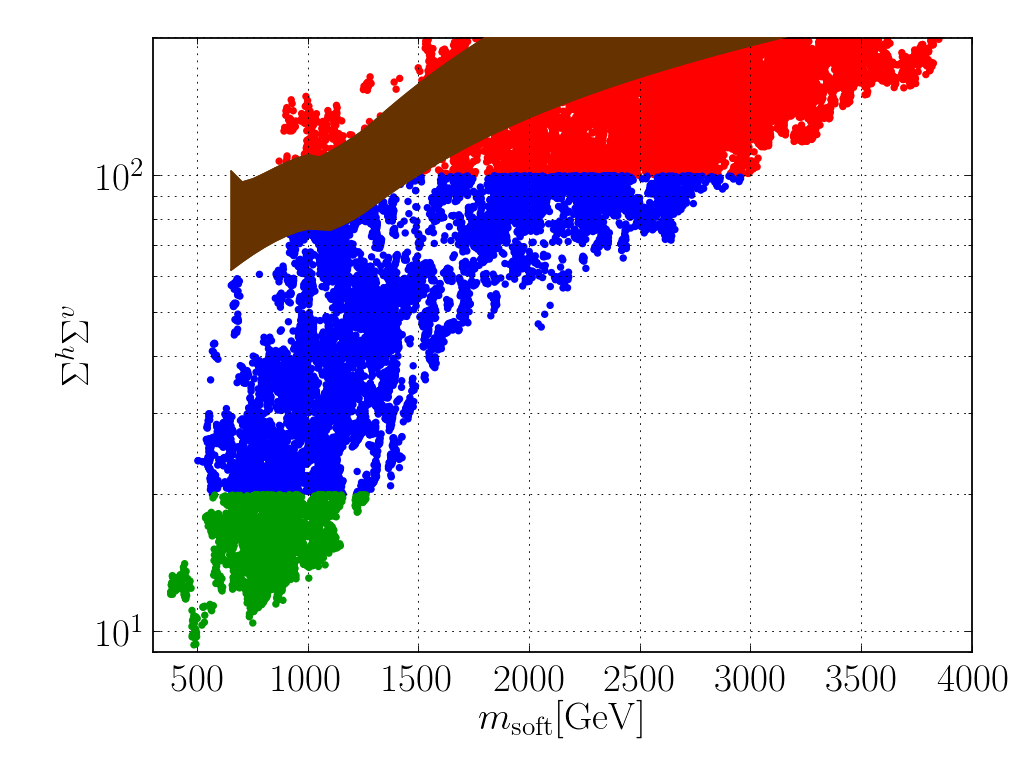}
\caption{\label{fig:totFTvsMSUSY}}
\end{subfigure}
\caption{\em The combined tuning $\Sigma^h\Sigma^v $ as a function of (a) the Higgs-singlet coupling $\lambda$ and (b) $\msoft$. The points in (b) have $\Mmess=20 \TeV$ and a tuning in the Higgs VEV better than $5\%$.  
The green, blue and red points in (b) correspond to a combined tuning better than $5\%$, between $1\%$ and $5\%$, 
and worse than $1\%$, respectively. The brown band in (b) corresponds to the Higgs VEV tuning for a generic region of the MSSM with $m_h\in [124,127] \GeV$.  All points satisfy the constraints discussed in \Secref{sec:collider}.
}
\label{fig:totaltuning}
\end{figure}

In \Figref{fig:totFTvslambda}, we show a scatter plot of the combined tuning as a function of $\lambda$. In addition to the messenger scale $\Mmess=20 \TeV$ mainly considered in this paper (black points), we have included points with $\Mmess=100 \TeV$ (orange) and $\Mmess=1000 \TeV$ (yellow). Note that the combined tuning increases with growing messenger scale. This is expected since loop corrections to soft parameters from RG running also increase with the messenger scale (cf.~Eq.~\eqref{mhutuning}). Furthermore, notice that there is a minimum for the tuning at values $\lambda \approx 1$. This can be understood as follows: Let us first consider the region $\lambda \gtrsim 1$. We expect that the smallest tuning measure $\Sigma^v$ which is achievable for a given $\lambda$, decreases with increasing $\lambda$ in this region. This is indeed what we observe in our sample of points. The trend, however, is less pronounced than the $\lambda^{-2}$-dependence that we may naively expect from the arguments in the preceding section.\footnote{We note that this partly results from our linear scanning of the parameter space (see \Secref{sec:NumAnal}) which prefers larger values for the soft masses and thereby counteracts the suppression of $\Sigma^v$ for large $\lambda$.} Since $\Sigma^h$ grows approximately like $\lambda^2$, on the other hand, this explains the growth with $\lambda$ of the combined tuning in the region $\lambda \gtrsim 1$ of the scatter plot. In the region $\lambda \lesssim 1$, the tuning measure $\Sigma^h$ becomes approximately independent of $\lambda$ as can be seen in \Figref{fig:higgstuning}. This corresponds to the fact that the Higgs mass at tree-level is below 126 GeV for $\lambda \lesssim 1$ so that no tuning arises in the Higgs mass. Since loop corrections from the top-stop sector are then required to raise the mass, the tuning measure $\Sigma^v$ in turn grows for decreasing $\lambda$. Together this explains the behavior of the combined tuning in the region $\lambda \lesssim 1$ of the scatter plot. 

In \Figref{fig:totFTvsMSUSY}, we show a scatter plot of the combined tuning as a function of $m_\soft = \sqrt{m_{Q_3} m_{u_3}}$. The brown band maps the tuning in the MSSM, where we have chosen $\Mmess= 20 \TeV$, $\mu=200 \GeV$, $\tan\beta=20$ and $m_a=1 \TeV$ for the mass of the $CP$-odd scalar to represent a generic region of parameter space. Furthermore, the $A$-term $A_t$ is varied such that ${m_h=124 \, (127) \GeV}$ for the lower (upper) boundary. We have used the program {\it FeynHiggs} \cite{Heinemeyer:1998yj} to calculate the Higgs mass. Note that there is no tuning in the Higgs mass required in the MSSM. Indeed, the mass at tree-level is always too small and large loop corrections are required to lift the mass to $126 \GeV$. We have accordingly set $\Sigma^h=1$ for the MSSM. We see that, even taking the combined tuning into account, there is a large region of parameter space for which the (scale-invariant) NMSSM requires less tuning than the MSSM. This is not surprising as values $\lambda \sim 1$ do not introduce a significant tuning in the Higgs mass but alleviate the need for heavy stops to raise the mass as in the MSSM.

There can also be a tuning associated with the lightest $CP$-odd Higgs state.  This can be seen by noting that 
the lightest mass-squared eigenvalue in (\ref{CPoddmatrix}) is given by
\begin{equation}
 m_{a_1}^2\simeq -3 a_\kappa v_S + 9 \frac{\lambda \kappa v^2}{m_A^2}a_\lambda v_S ~.
\end{equation}
Clearly if  $m_{a_1}\lesssim 100$ GeV with soft mass parameters in the TeV range, a tuning will occur, which was
verified numerically: a tuning in $m_{a_{1}}$ at the 5$\%$ level is found for $m_{a_{1}}\lesssim 100$ GeV for parts of the parameter space. We also checked numerically that all of the other particle and sparticle masses in the regions of interest have tunings better than 50$\%$. So we conclude that if at some time in the future a $CP$-odd scalar is discovered with mass $m_{a_{1}}\lesssim 100$ GeV it could result in an additional source of tuning from the perspective of the model under consideration.



\section{Limits from Colliders and Cosmology}
\label{sec:collider}
Physics beyond the SM is already strongly constrained by direct searches at colliders, as well as indirectly by 
electroweak precision tests and flavor physics. Several such limits (in particular from LEP, the Tevatron and the B-factories) are already implemented in {\it NMSSMTools} 3.2.1. We apply all these limits in our scan, except for the flavor constraints for which we use updated results. Also note that we do not attempt to explain the discrepancy in the measurement of the anomalous magnetic moment of the muon, $(g-2)_{\mu}$, with the SM expectation. Furthermore, we require that the LSP abundance does not overclose the universe and that it satisfies direct detection limits.  
 
In addition to the limits implemented in {\it NMSSMTools} 3.2.1, we take into account several searches at the LHC as well as electroweak precision tests. In the following, we focus on our additional constraints and refer the interested reader to the documentation of {\it NMSSMTools} 3.2.1 for the remaining limits.

\subsection{Direct limits from colliders}
\label{sec:colliderlimits}

We identify the bosonic resonance, which was recently discovered in the diphoton and diboson channels at ATLAS \cite{:2012gk} and CMS \cite{:2012gu}, with one of the three  $CP$-even states in the Higgs sector of the NMSSM. We denote this state by $h$. In agreement with the current experimental uncertainties, we require that its mass lies in the range ${124 \GeV \lesssim m_{h}\lesssim 127 \GeV}$. Furthermore, we require that its signal strength is within the 1$\sigma$ range measured at the LHC for the most constraining channels. More precisely, we take the average value of the signal strengths quoted by ATLAS~\cite{:2012gk,ATLAS-CONF-2012-162,ATLAS-CONF-2012-170} and CMS~\cite{:2012gu,CMS-HIG-12-045} for each channel and combine the errors in quadrature. Here the signal strength is defined as
\begin{equation}
R_{X}\, \equiv\, \frac{\sigma(h) \times BR(h\to X)}{\sigma(h_{\rm SM}) \times BR(h_{\rm SM} \to X)}\,,
\end{equation}
where $h_{\rm SM}$ denotes the SM Higgs, $\sigma$ the production cross section at the LHC and $BR$ the branching fraction to final states $X$. We then require that
\begin{align}
0.81<R_{ZZ}<1.32, \qquad \, &
0.74<R_{WW}<1.40  , \nonumber \\
0 <R_{b\bar b }<1.10 , \qquad \, &
0.27 <R_{\tau\tau}<1.15  ,
\end{align}
where for $R_{b\bar b }$, the Higgs production is assumed to be in association with a vector boson.  However, in order to allow for a Higgs with SM properties in our scan, we do not impose the corresponding range for the diphoton channel:
\begin{equation}\label{eq:Rgammagamma}
1.40<R_{\gamma\gamma}<1.99 \, .
\end{equation}
Instead, we discuss it separately in \Secref{sec:pheno}.

The Higgs searches by ATLAS~\cite{:2012gk} and CMS~\cite{:2012gu} also constrain the heavier $CP$-even states $s_2$ and $s_3$. At high masses, decays into $WW$ and  $ZZ$ dominate. These channels are highly constrained and the production cross section times branching fraction (into $WW$ and $ZZ$) of these states has to be significantly smaller than the corresponding rate of a SM Higgs at the same mass~\cite{CMS-PAS-HIG-11-024,CMS-PAS-HIG-12-041}. More precisely, we shall require that
\begin{equation}
\frac{\sigma(s_{i}) \times BR(s_{i}\to ZZ)}{\sigma(h_{\rm SM}) \times BR(h_{\rm SM}\to ZZ)} <0.09 , \qquad
\frac{\sigma(s_{i}) \times BR(s_{i}\to WW)}{\sigma(h_{\rm SM}) \times BR(h_{\rm SM}\to WW)} <0.2 .
\end{equation}

Besides limits on $CP$-even Higgs bosons, we impose all bounds on sparticle masses coming from LEP and the Tevatron. In particular, these bounds are (cf.~\cite{PDG:2012}) $m_{\tilde t}>95.7\GeV$ for the stop, $m_{\tilde b}>89 \GeV$ for the sbottom, $m_{H^\pm}\gtrsim79.3\GeV$ for a charged Higgs, and $m_{\tilde \chi^\pm}\gtrsim94\GeV$ for a chargino.  

Furthermore, we take into account the latest SUSY searches at the LHC. The ATLAS and CMS experiments have presented exclusion plots for sparticle masses in various simplified models, where most of the sparticles are decoupled. Note that, for a typical point in our parameter space, more sparticles will be light
than assumed in these simplified models. This can lead to longer decay chains as those considered in these studies and could weaken certain limits or make them more stringent. In our scan, we shall nevertheless impose bounds derived from these simplified models. We do not expect that our results are affected by this simplified treatment of the LHC limits for the following reasons: Firstly, we impose bounds which are more conservative than those coming from the simplified models (the precise limits will be given below). This leaves some room in case the simplified-model limits are not sufficient to ensure that a given point in our parameter space does not lead to a visible excess in the LHC searches. Secondly, 
we are mostly interested in the upper limits on sparticle masses coming from naturalness, whereas direct searches give lower limits. Therefore, we expect that our results do not depend sensitively on the latter.

The experimental limits depend on the nature of the LSP.  Let us therefore first discuss the different possibilities for the LSP in our scenario. 
In order to minimize the size of loop corrections to soft masses, we require that soft breaking terms are generated at a low scale $\Mmess$ (which we refer to as the `messenger scale') . One can imagine various types of UV completions above this scale. 
For example, $\Mmess$ could be associated with the mass of messenger fields like in gauge-mediated SUSY breaking. In this case, a gravitino with mass $\smash{m_{3/2} \approx \msoft \Mmess / M_{\rm Pl} \approx 0.01 \eV}$ would be the LSP. Alternatively, $\Mmess$ could be a cutoff at which the Higgs sector and/or other NMSSM fields emerge as composites of an underlying strongly coupled theory. This is, for example, realized in the models of \cite{Gherghetta:2003he,Harnik:2003rs,Sundrum:2009gv,Gherghetta:2011wc}. This case allows for the possibility that the LSP resides in the visible sector instead of being the gravitino. Early-universe cosmology then favors a neutralino as the LSP. 

In our scan, we shall focus on the case of a neutralino LSP. This is partly motivated by our desire to study the prospects of dark matter in this scenario (gravitinos with mass $\smash{m_{3/2} \approx 0.01 \eV}$ are underabundant). In addition, there are currently only a few searches which consider a spectrum with a gravitino LSP (usually in the framework of general gauge mediation) and decoupled first-two-generation sparticles. This makes the implementation of LHC constraints without performing detailed simulations difficult. 

We believe, however, that our results are relatively independent of this choice of the LSP. This is for the following reasons: We have performed a dedicated scan in which we made no requirements on the nature of the lightest superpartner (i.e.~the NLSP if the gravitino is the LSP). After imposing the limits discussed so far in this section and those from LEP and the Tevatron, we have found that in 98\% of the cases the LSP is a neutralino (with various admixtures of wino, bino, singlino and Higgsino components). Requiring that the LSP is a neutralino, as we do in our scan, thus does not eliminate many points. There are dedicated searches for gluinos and stops in scenarios with a gravitino LSP, decoupled first-two-generation sparticles and a bino-like neutralino NLSP \cite{ATLAS-CONF-2012-072,Aad:2012cz,Barnard:2012au} (in events with diphotons plus \met) or a Higgsino-like neutralino NLSP \cite{Aad:2012cz} (in events with a Z boson, jets plus \met). The limits on stops and gluinos masses are comparable to those for a neutralino LSP except that the constraints are almost independent of the mass of the neutralino NLSP (in particular $m_{\tilde{g}}\lesssim 1.3 \TeV$ is excluded for all NLSP masses $m_{\tilde{\chi}^0_1}\lesssim m_{\tilde{g}}$).
We therefore expect that the set of points which satisfy the current LHC limits does not differ significantly for the cases of a gravitino and neutralino LSP. Finally, dark matter constraints eliminate some points for a neutralino LSP (due to overclosure and direct detection limits as discussed in \Secref{DMsection}). No such constraints apply for a gravitino LSP with mass $\smash{m_{3/2} \approx 0.01 \eV}$ as its abundance is sufficiently  small. However, we have found that less than 1\% of the points which survive all other limits are eliminated by the dark matter constraints in the former case.
For these reasons, we expect that the set of points which pass all constraints would not change significantly if we considered a gravitino LSP instead of a neutralino LSP. 

Let us now discuss the LHC limits that we impose for the case of a neutralino LSP. An important search by ATLAS uses final states with b-jets and missing transverse energy (\met) in 12.8 fb$^{-1}$ of data taken at 8 \TeV~\cite{ATLAS-CONF-2012-145}. Bounds are derived in two simplified models where gluinos are directly produced and decay to the neutralino LSP and either $b\bar{b}$ or $t\bar{t}$ (with 100\% branching fraction). All sparticles except the gluino and the neutralino are decoupled. Since the limits from the decay $\tilde g \to b\bar b\tilde\chi^0_1$ are slightly more stringent than those from $\tilde g \to t\bar t\tilde\chi^0_1$, we will impose the former. Following the exclusion curve presented  in \cite{ATLAS-CONF-2012-145} for this case, we shall conservatively exclude the region
\begin{equation}\label{eq:GluinoBound}
m_{\tilde g}<1310\GeV \qquad \text{if}\;\; m_{\tilde\chi^0_1}<650\GeV \, .
\end{equation}
Another search by ATLAS in events with jets, leptons and \met~\cite{ATLAS-CONF-2012-151} gives weaker constraints on the gluino mass. Furthermore, CMS has used less data for its gluino searches than ATLAS and the resulting limits are correspondingly less stringent too.

An important search by CMS uses final states with b-jets and $\met$ in 11.7 fb$^{-1}$ of data at $8\TeV$  \cite{SUS-12-028-pas}. Constraints are derived on a simplified model, where the sbottom is directly produced and decays with 100\% branching fraction into the neutralino LSP and a b-quark. Again all other sparticles are decoupled. Following the relevant exclusion curve presented in \cite{ATLAS-CONF-2012-106,SUS-12-028-pas}, we will forbid points with
\begin{equation}\label{eq:SbottomBound}
150 \GeV<m_{\tilde b} < 650\GeV \qquad \text{if $m_{\tilde\chi^0_1}<230\GeV$} \, .
\end{equation}
Note that ATLAS did not update their sbottom search using 8 $\TeV$ data yet and their bounds are correspondingly weaker. An analogous simplified model with stops instead of sbottoms is constrained by two ATLAS searches: In \cite{1208.1447}, final states with jets and \met are analyzed using 4.7 fb$^{-1}$ of data at 7 $\TeV$. In \cite{1208.2590}, these final states are required to also have an isolated lepton. Following the exclusion plots in these papers, we shall conservatively forbid points in the region 
\begin{equation}\label{eq:StopBound}
220 \GeV<m_{\tilde t}<500\GeV\qquad \text{if}\;\; m_{\tilde\chi^0_1}<160\GeV \, .
\end{equation}
Note that, even though the latest stop search by CMS  \cite{CMS-PAS-SUS-12-023} uses 8 $\TeV$ data and higher integrated luminosity, it results in comparable limits. 

Final states with leptons and \met are used in dedicated searches for neutralinos and charginos by ATLAS \cite{ATLAS-CONF-2012-154} and CMS \cite{Chatrchyan:2012ewa,CMS-PAS-SUS-12-022}. Limits are derived in a simplified model where the chargino-neutralino pair $\tilde{\chi}_1^{\pm} \tilde{\chi}_2^{0}$ is produced and decays with 100\% branching fraction to $\tilde{\chi}_1^{0} Z \tilde{\chi}_1^{0} W^{\pm}$. Furthermore, to simplify the parameter space, it is assumed that the LSP $\tilde{\chi}_1^{0}$ is bino-like, $\tilde{\chi}_1^{\pm}$and $\tilde{\chi}_2^{0}$ are wino-like and that the latter two particles have the same mass. The more stringent limit comes from the CMS search \cite{CMS-PAS-SUS-12-022} which excludes approximately the region
\begin{equation}
m_{\tilde{\chi}_1^{\pm}}  = m_{\tilde{\chi}_2^{0}} < 330 \GeV \qquad  \text{if $m_{\tilde{\chi}_1^{0}} < 120 \GeV \, $ and $ \, m_{\tilde{\chi}_1^0} + m_W < m_{\tilde{\chi}_1^{\pm}}$} \, .
\end{equation}
The limit does not apply for  $\smash{m_{\tilde{\chi}_1^0} + m_W > m_{\tilde{\chi}_1^{\pm}}}$ because on-shell $W$s and $Z$s are assumed in the CMS search.
We find that this does not pose any constraint on our generated data points, because there are only a few points with a dominantly bino LSP and $\tilde{\chi}_1^{\pm},\tilde{\chi}_2^{0}$ being wino-like. All of these points have an LSP mass $m_{\tilde\chi_1^0}>120\GeV$.

\subsection{Electroweak Precision Tests}
\label{sec:EWPT}

The additional particles of the NMSSM compared to the SM contribute to the Peskin-Takeuchi parameters~\cite{Peskin:1991sw} $S$ and $T$.  
Using the latest measurements,
the Particle Data Group~\cite{PDG:2012} quotes the ranges $S_0=-0.04\pm0.09$ and $T_0=0.07\pm 0.08$ at $95\%$ C.L. for a reference Higgs mass of $m_{h,ref}=117$ GeV with a correlation between $S$ and $T$ of 88$\%$. We constrain the model parameters by demanding accordance with these measurements at $95\%$ C.L. .

We can distinguish three sectors that contribute to $S$ and $T$: i) Higgs and singlet scalars, ii) stops and sbottoms and iii) neutralinos and charginos. 
Since the expressions for the contributions of these sectors can be found in the literature, we shall not repeat them here. We use the formulas given in~\cite{Cho:1999km, Hagiwara:1994pw, Martin:2004id} for the neutralino-chargino and stop-sbottom sector and those from \cite{Barbieri:2006bg,  Franceschini:2010qz} for the singlet and Higgs scalars. In order to compare with the values quoted by the Particle Data Group, the contribution of a SM Higgs with reference mass $m_{h,ref}=117$ GeV has to be subtracted from the latter.  

Note that we do not decouple the electroweak gauginos but take all charginos and neutralinos into account, in contrast to the analysis in \cite{Barbieri:2006bg, Franceschini:2010qz}.  
In these references, it was found that accordance with electroweak precision tests requires $\tan\beta \lesssim 3$ for large $\lambda \approx 2$. Here the largest contributions come from the stop-sbottom and chargino-neutralino sector. We have verified that we recover a similar constraint on $\tan \beta$ in the limit of decoupled electroweak gauginos and $\lambda\approx 2$.   
Smaller $\lambda$ allows for larger $\tan\beta$ and in our scan, we find values up to $\tan\beta\approx 4.2$ for $\lambda\approx 1.1$.
This dependence on $\lambda$ is related to the breaking of the custodial SU(2) for $\tan\beta\neq 1$ (which in turn leads to a correction to the $T$ parameter). Indeed, the Higgsino-singlino sector can provide sizable contributions to the $T$ parameter due to large off-diagonal terms in the mass matrix (see \cite{Franceschini:2010qz} for a discussion of this in the limit of decoupled electroweak gauginos). This mixing and thus the effect of custodial symmetry breaking due to $\tan\beta\gtrsim 1$ is enhanced for large $\lambda$.  

\subsection{Flavor Constraints}
\label{sec:FlavorConstraints}
The package {\it NMSSMTools} allows to calculate several flavor observables. We use the calculated values to impose the current constraints.
For the mass differences in the $B^0$ system,  $\Delta M_d$, and the $B^0_s$ system, $\Delta M_s$, we use the current $2\sigma$ ranges from the HFAG group~\cite{Amhis:2012bh}:
\begin{align*}
& \Delta M_s =(17.719\pm0.086)\, \mathrm{ps}^{-1} \,,\\
& \Delta M_d =(0.507\pm0.008)\, \mathrm{ps}^{-1}\, .
\end{align*}
We also impose the $2\sigma$ ranges on the branching ratios of the $B$ decays $B^+\to\tau^+\nu_\tau$ and $B\to X_s\gamma$ from the same group:
\begin{align*}
& \mathrm{Br}(B^+\to\tau^+\nu_\tau) =(1.67\pm0.60)\times 10^{-4}\,, \\
& \mathrm{Br}(B\to X_s\gamma) =(3.55\pm0.48\pm0.18)\times 10^{-4}\;.
\end{align*}
In case of the rare decay $B_s^0\to\mu^+\mu^-$, we use the $2\sigma$ ranges from the recent LHCb measurement~\cite{arxiv-1211.2674}:
\begin{equation*}
\mathrm{Br}(B_s^0\to\mu^+\mu^-)=3.2^{+3.0+1.0}_{-2.4-0.6}\times 10^{-9}\;.
\end{equation*}

Since small values of $\tan\beta\sim 1$ are needed in order to have a sizable tree-level contribution to the Higgs mass and to satisfy electroweak precision tests in our model,
the dominant flavor constraints arise from the contribution of the charged Higgs to $B\to X_s\gamma$.  Indeed, this contribution dominates in the small $\tan\beta$ limit 
when $\mu A_t>0$, while other flavor changing effects become only relevant when $\tan\beta\gg 1$~\cite{Domingo:2007dx}.

We do not attempt to explain the discrepancy in the measurement of the anomalous magnetic moment of the muon, $(g-2)_{\mu}$, with the SM expectation.  Furthermore, we do not consider any constraints from lepton flavor violating processes since we assume decoupled sleptons, $m_{\tilde{l}}\approx \Mmess$.

\subsection{Cosmological Constraints and Dark Matter}
\label{DMsection}

As discussed in \Secref{sec:colliderlimits}, we focus on the case of a neutralino LSP in our scan. Since we assume $R$-parity conservation, we have to ensure that the LSP relic abundance does not overclose the universe. Using the WMAP-7 measurements of the dark matter abundance~\cite{Larson:2010gs}, we require that $\Omega_{\rm LSP} h^2\leq \Omega_{\rm WMAP-7} h^2= 0.1120\pm 0.0056$. In addition, we impose the latest limits on direct detection from XENON100 \cite{Aprile:2012nq}. We calculate  the relic abundance and the direct-detection cross sections using the program  {\it MicrOMEGAs} \cite{Belanger:2005kh,Belanger:2010gh}.

Note, however, that constraints from overclosure and direct detection are avoided if the gravitino is the LSP. Indeed, if the scale $\Mmess$ is associated with the mass of messenger fields like in gauge mediation, we expect the gravitino to have a mass $m_{3/2}\approx 0.01 \eV$ (see \Secref{sec:colliderlimits}). The goldstino component of the gravitino couples with strength  $1/(m_{3/2}M_{P})$, where $M_{P}$ is the reduced Planck mass. Such a light gravitino is therefore in thermal equilibrium with the plasma in the early universe~\cite{deGouvea:1997tn}. After freeze-out, its relic abundance is $ \Omega_{3/2} h^2\approx (m_{3/2}/{\rm keV})\times (100/g^{*}(T_f))$, where $g^{*}(T_f)$ is the effective number of degrees of freedom at freeze-out (typically $g^{*} \sim 100-200$)~\cite{Giudice:1998bp}. For $m_{3/2}\approx 0.01$ eV, we thus find that $\Omega_{3/2} h^2\ll \Omega_{\rm WMAP-7} h^2$.  This also avoids all constraints from direct detection.
Furthermore, we note that the available collider limits for a scenario with neutralino NLSP and gravitino LSP are not significantly different from the limits for a neutralino LSP (see \Secref{sec:colliderlimits}). In this section, we shall therefore include points which would be excluded for a neutralino LSP and will interpret them in the context of a scenario with neutralino NLSP and gravitino LSP.\footnote{Alternatively, these points could be included if the reheating temperature is sufficiently low (see e.g.~\cite{Fornengo:2002db}).}
 In particular, we require the gluino for these points to satisfy the limit ${m_{\tilde{g}}\gtrsim 1.3 \TeV}$ (cf.~\Secref{sec:collider}).

\begin{figure}[bt]\centering
\includegraphics[width=0.7\linewidth]{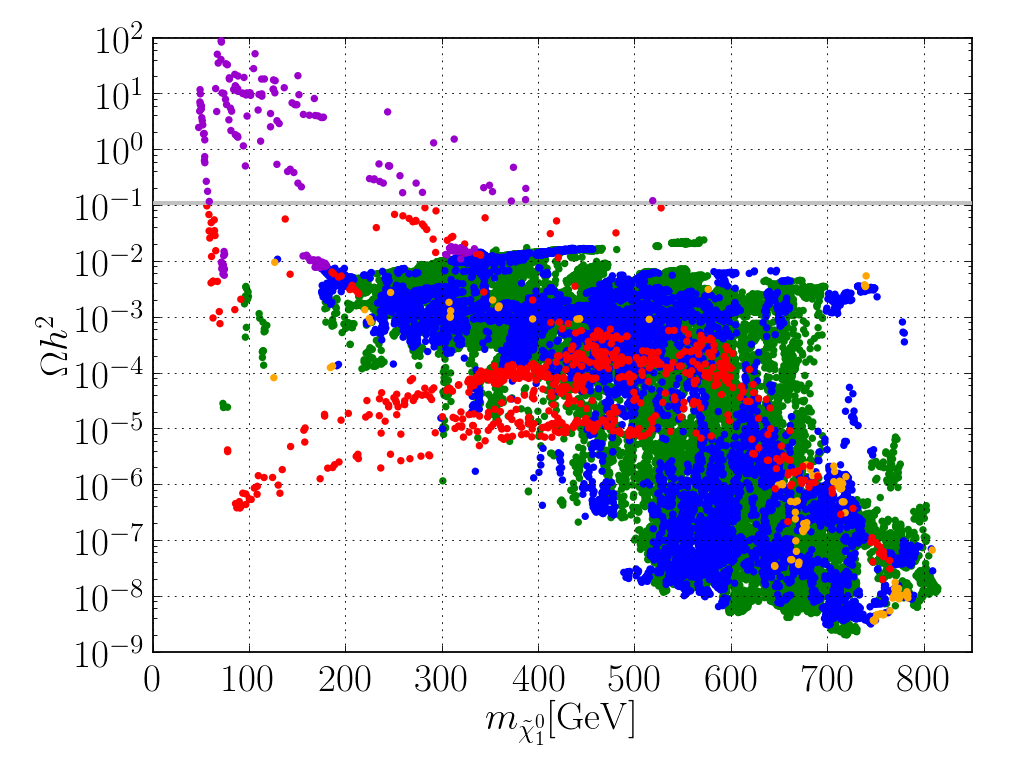}
\caption{\em A log-linear plot of the LSP relic density $\Omega h^2$ versus the LSP mass $m_{\tilde{\chi}^{0}_1}$.  The colors denote the dominating component of the neutralino: green, blue, orange, red corresponds to a lightest neutralino which is dominantly singlino, Higgsino, wino or bino, respectively. The purple points are excluded by the WMAP-7 or XENON100 constraints, but they are still viable for a scenario with gravitino LSP and neutralino NLSP. The gray band depicts the experimentally measured DM relic density. All points satisfy the constraints discussed in sects.~\ref{sec:colliderlimits} to \ref{sec:FlavorConstraints}.}
\label{fig:RelicDensiy}
\end{figure}

The scatter plot in Fig.~\ref{fig:RelicDensiy} shows the relic density $\Omega h^2$ of the neutralino LSP as a function of its mass. We have colored the points according to the species which dominates the neutralino composition: green, blue, orange, and red  points correspond to neutralinos which are respectively mostly singlino, Higgsino, wino or bino.  Purple points, on the other hand, are excluded due to overclosure or direct detection for a neutralino LSP and accordingly require a gravitino LSP (in which case the neutralino NLSPs decay on short time scales to gravitinos and SM particles). It turns out that the neutralino NLSP for these points is mostly bino.

Note that for most of our points, the neutralino LSP is underproduced compared to the relic density which is required to account for the dark matter. This meets our expectations for a neutralino which is mostly wino or Higgsino, since its couplings to the $Z$-boson and the Higgs then yield a large annihilation cross section.  It was found in \cite{Barbieri:2006bg,Franceschini:2010qz}, on the other hand, that the right relic abundance can be obtained for a mostly-singlino neutralino. It is possible that our scan missed such points because of the scanning procedure or the fine-tuning constraint. We also note that ref.~\cite{Barbieri:2006bg} considered a superpotential which is at most quadratic in the singlet superfield $S$ (ref.~\cite{Franceschini:2010qz} assumed the scale-invariant NMSSM though). The presence of the cubic term in the scale-invariant NMSSM allows for new annihilation channels and results thus in a smaller relic abundance for a mostly-singlino neutralino. We find that the $CP$-even scalar $s_2$ (the second-lightest after the Higgs) is typically mostly singlet and not very heavy, $m_{s_2}\lesssim 1.1$ TeV.  The cubic interaction in the superpotential then allows for strong s-channel annihilation of a mostly-singlino neutralino via $s_2$. Similarly, if the Higgs has a significant singlet admixture, the Higgs can mediate the s-channel annihilation. 

\begin{figure}\centering
\begin{subfigure}{0.3\linewidth}
\FMDG{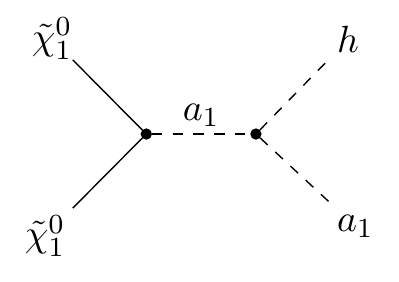}
\caption{}
\end{subfigure}
\hfill
\begin{subfigure}{0.3\linewidth}
\FMDG{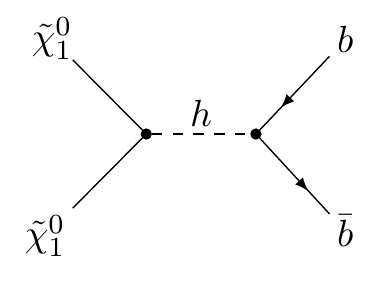}
\caption{}
\end{subfigure}
\hfill
\begin{subfigure}{0.3\linewidth}
\FMDG{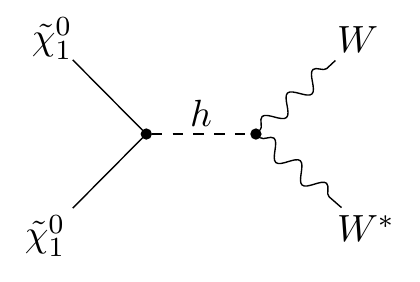}
\caption{}
\end{subfigure}
\caption{\em Main annihilation  channels of the neutralino LSP for points with relic density ${\Omega_{\tilde{\chi}^0_1}h^2\approx 0.1}$. Though $\tilde{\chi}^0_1$ is mostly bino for these points, its Higgsino component provides the strongest annihilation channels. }
\label{fig:DMannihilation}
\end{figure}

We find a few points for which the right relic density is obtained to account for the dark matter. For these points, the neutralino LSP is mostly bino with a small admixture of wino and Higgsino, whereas the singlino component is negligible. Its mass lies in the range $m_{Z}/2 \lesssim m_{\tilde{\chi}^0_1}\lesssim 500 \GeV$ (the lower limit results from the constraint on the invisible decay width of the $Z$). Furthermore, we find that the lightest chargino is mostly Higgsino with mass $m_{\tilde{\chi}^{\pm}_1}\gtrsim 200 \GeV$, whereas the charged Higgs has a mass $m_{H^{\pm}}\gtrsim 750$ GeV. We therefore expect that  t-channel annihilation of the LSP via a chargino and LSP-chargino co-annihilation  is suppressed due to the heavy charged Higgs. The second-lightest neutralino, on the other hand, is mostly singlino with a small wino component and a mass $m_{\tilde{\chi}^0_2}\gtrsim 350 \GeV$. This makes significant co-annihilation of this state with the LSP unlikely. Stops and sbottoms also have large masses, $m_{\tilde{t}_1, \tilde{b}_{1}}\gtrsim 1 \TeV$,  which suppresses annihilation channels involving squarks. Instead, we find that the most relevant annihilation channels result from the small but non-vanishing Higgsino component of the neutralino LSP. This allows for s-channel annihilation via $a_1$ to $h  a_1$ or via the Higgs to $b \bar{b}, W W$, where the Higgs is typically on resonance, $2m_{\tilde{\chi}^0_1}\approx m_h$.  The corresponding Feynman diagrams are shown in \Figref{fig:DMannihilation}. The fact that the Higgsino component of the neutralino LSP is rather small is compensated by the large $\lambda$-coupling. To recapitulate, we find that 
the right relic density can be obtained for a neutralino LSP if it is mostly bino and has a small but non-vanishing Higgsino component.

\section{Phenomenological Implications}
\label{sec:pheno}

\subsection{Particle spectrum}

The spectrum of particles in the scale-invariant NMSSM that plays a dominant role in the Higgs sector is constrained by two competing effects. Naturalness prefers the Higgsino, stop/sbottom and gluino to be light, while experimental constraints, mostly from the 8 TeV LHC run, place increasingly sizable lower limits on these sparticle masses. Nevertheless there still remains a range of particle masses that has not yet been fully explored at the LHC where the
tuning is not too severe.

\begin{figure}[htb]\centering
\includegraphics[width=15cm]{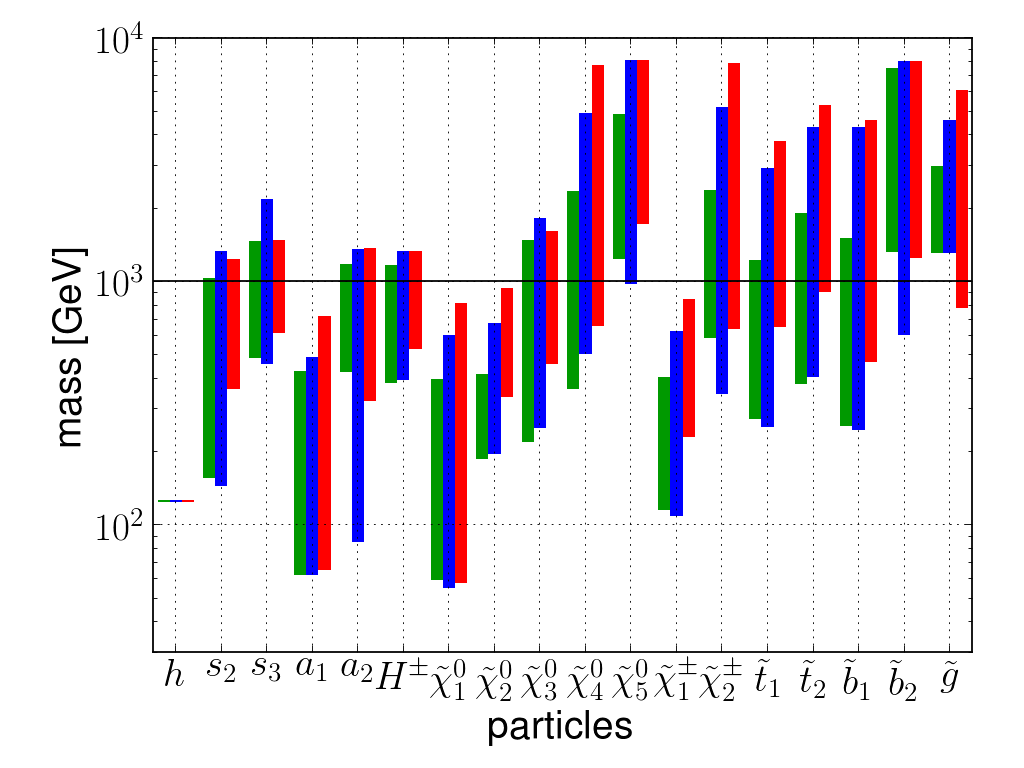}
\caption{{\em The particle spectrum of the scale-invariant NMSSM with $\Mmess=20\TeV$ that we find in our scan satisfying all constraints described in the text. The green, blue and red points correspond to a combined tuning ($\Sigma^h\Sigma^v$) better than $5\%$, between $1\%$ and $5\%$, and worse than $1\%$, respectively.
All points satisfy the constraints discussed in \Secref{sec:collider}.
Note that there are correlations among the masses which are not shown in the figure, but are described in \Secref{sec:collider}. }}
\label{fig:spectrum}
\end{figure}

In Fig.~\ref{fig:spectrum}, we show the ranges of masses in our scan that satisfy all constraints described in sections \ref{sec:NumAnal}, \ref{naturalness} and \ref{sec:collider}. An effective $\mu$-term in the range $\mu\in[260,840]$ GeV is generated.  The state $h$ in the mass window [124,127] GeV (that we identify with the newly discovered bosonic resonance at the LHC) always corresponds to the lightest $CP$-even state. If we demand a total tuning better than 1$\%$, this state is mostly $h_u$ in its composition.  Notice that usually independently of  tuning,
most particles in the Higgs-singlet sector and part of the neutralino-chargino sector can have masses below 1 TeV.  Most prominently, we find such low masses for the states  $s_2$, $a_1$, $a_2$ and $\tilde{\chi}^0_1$,  $\tilde{\chi}^0_2$, $\tilde{\chi}^\pm_1$, respectively. In particular, the lightest chargino for a combined tuning better than 1$\%$ has a mass ${m_{\tilde{\chi}^{\pm}_{1}}\lesssim 575 \GeV}$ which can lead to loop-induced enhancement in the Higgs diphoton decay (cf.~\Secref{sec:Higgssearches}). The colored sector tends to be heavier but the possibility of stops, sbottoms and gluinos in the windows $m_{\tilde{t}_1}\in [220,1000]$ GeV, $m_{\tilde{b}_1}\in [216,1000]$ GeV, $m_{\tilde{g}}\gtrsim 1.2$ TeV remains viable.  In contrast to the MSSM however, the lightest stop is not necessarily light. We find stop masses $m_{\tilde{t}_1}$ up to $2.5 \TeV$ for a combined tuning better than 1$\%$. Given that in the absence of new colored states, the colorless states are only produced via electroweak processes, it will therefore require searches at the 14 TeV LHC to completely cover all natural regions of the spectrum. 
Notice that $m_{\tilde{\chi}^0_1}\gtrsim m_{Z}/2$  and $m_{a_1}\gtrsim m_h/2$ in the spectrum which follows from constraints on the invisible decay width of the $Z$-boson and the Higgs, respectively. 
Finally it is interesting to focus on the least-tuned region which is colored in green in the spectrum. As expected for small total tuning, the colored sector tends to be generically lighter though there also light colored sparticles in the more tuned regions (colored in red and blue). 

\subsection{Higgs searches at the LHC}
\label{sec:Higgssearches}
\begin{figure}[tb!]\centering
\begin{subfigure}{0.49\linewidth}
\includegraphics[width=\linewidth]{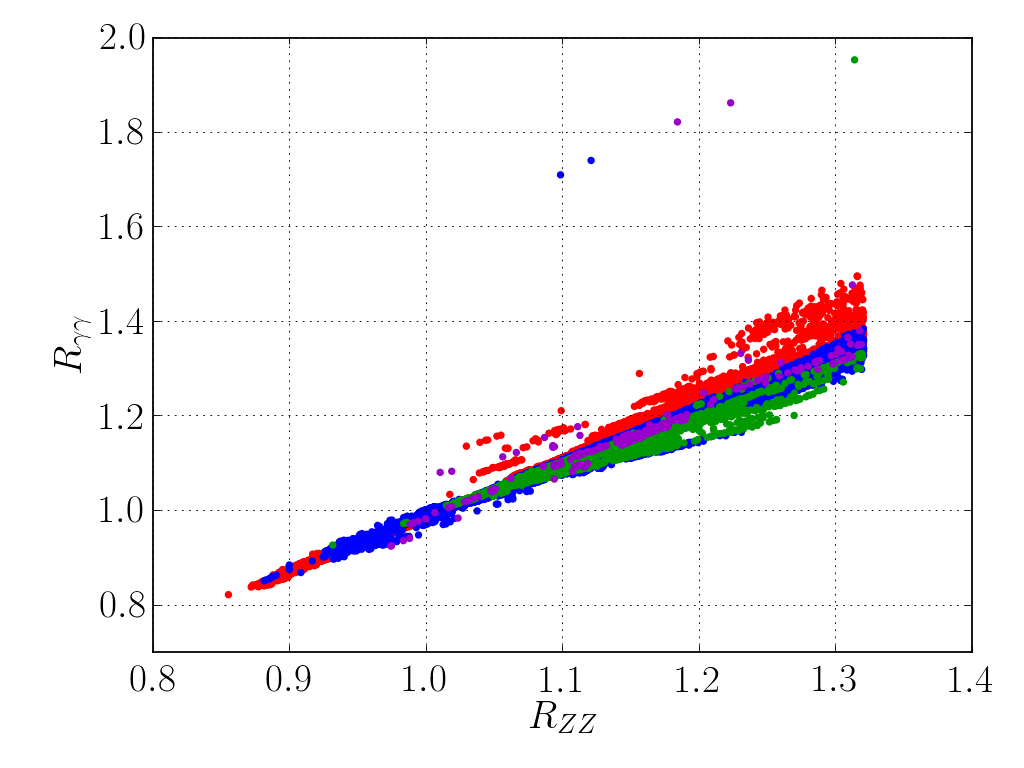}
\caption{\label{fig:ratesZZ}}
\end{subfigure}
\hfill
\begin{subfigure}{0.49\linewidth}
\includegraphics[width=\linewidth]{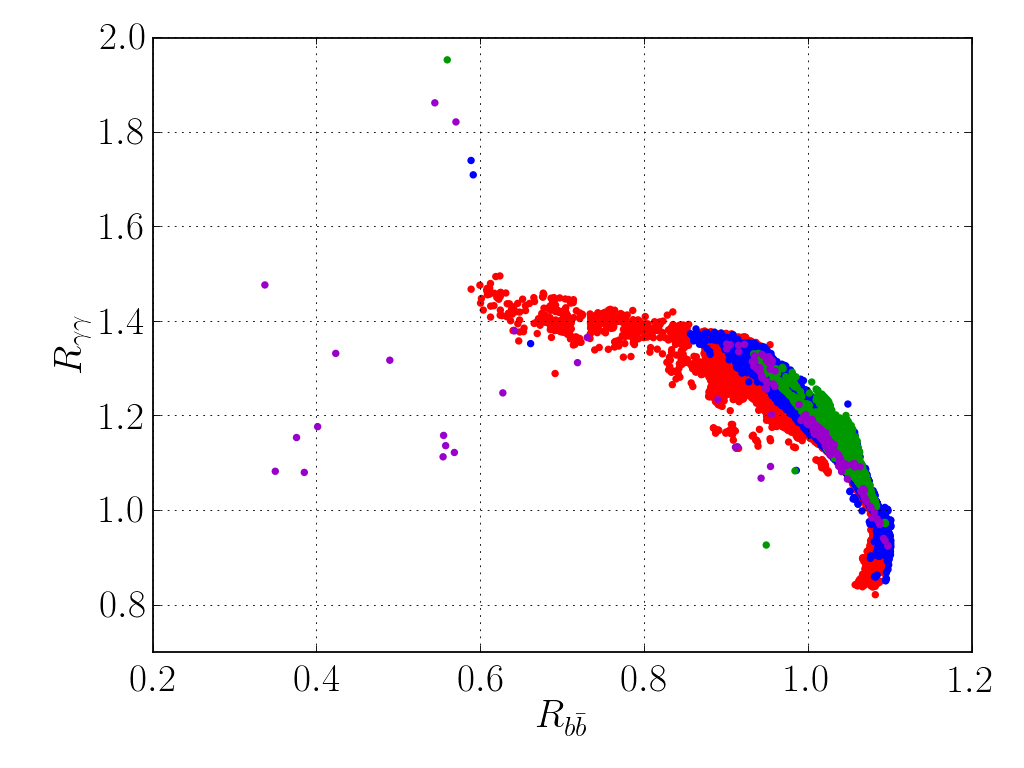}
\caption{\label{fig:ratesbb}}
\end{subfigure}
\begin{subfigure}{0.49\linewidth}
\includegraphics[width=\linewidth]{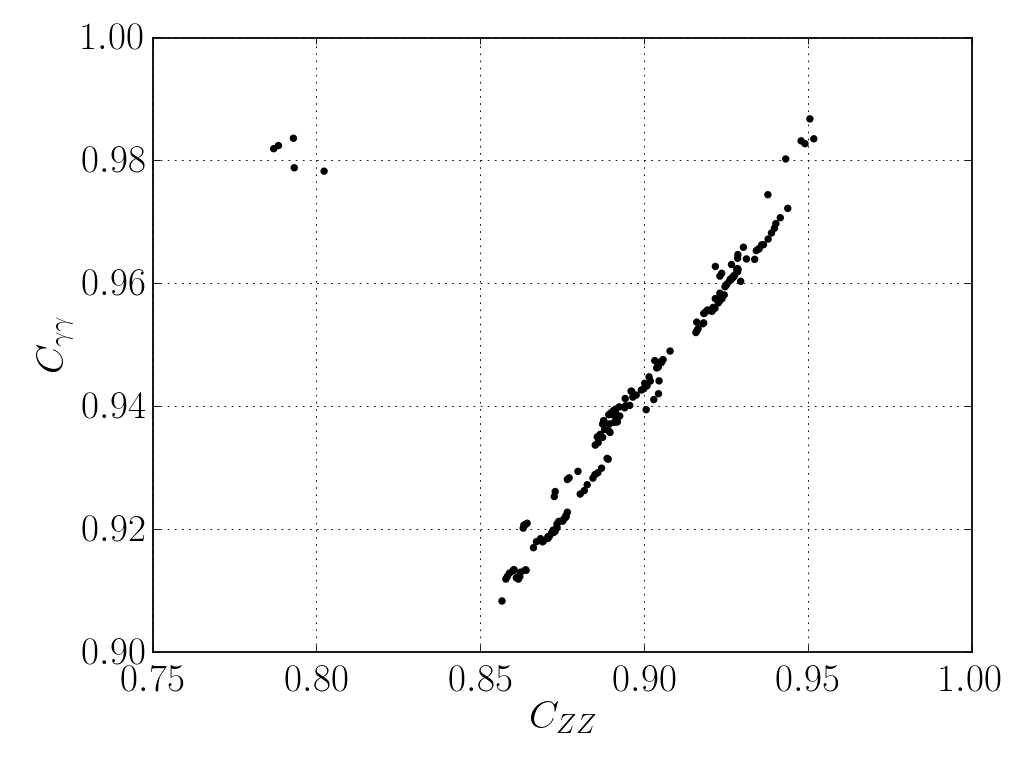}
\caption{\label{fig:enhanced2}}
\end{subfigure}
\hfill
\begin{subfigure}{0.49\linewidth}
\includegraphics[width=\linewidth]{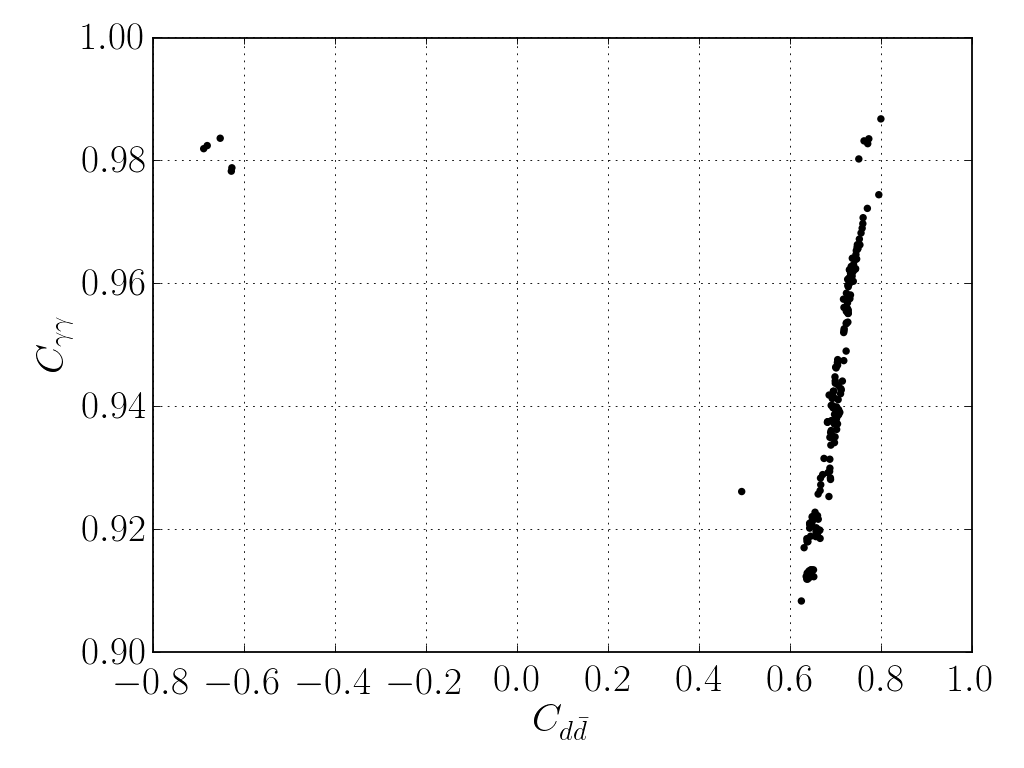}
\caption{\label{fig:enhanced1}}
\end{subfigure}
\caption{\em Scatter plots of the Higgs signal strengths $R_{ \gamma\gamma}$ versus (a) $R_{ZZ}$ and (b) $R_{ b\bar{b}}$. Here Higgs production is exclusively through gluon fusion except for the $b\bar{b}$ final state where production is in association with a vector boson. 
The green, blue and red points correspond to a combined tuning better than $5\%$, between $1\%$ and $5\%$, and worse than $1\%$, respectively.
The purple points correspond to a gravitino LSP and have a total tuning better than $1\%$. 
The scatter plots of the couplings $C_{\gamma\gamma}$ versus (c) $C_{ZZ}$ and (d) $C_{d\bar d}$ contain only points from the region with enhanced diphoton decay, $1.40<R_{\gamma\gamma}<1.99$ (cf.~\Eqref{eq:Rgammagamma}). Here the VEV fine-tuning is better than $5\%$. All points satisfy the constraints discussed in \Secref{sec:collider}.}
\label{fig:rates}
\end{figure}

In light of the recently discovered bosonic resonance at the LHC, the particle spectrum that we find leads to implications for the Higgs couplings and decays. In particular, we will consider decays into diphotons, dibosons ($WW^*$ and $ZZ^*$), $b\bar{b}$ in association with a vector boson and $\tau\bar{\tau}$ via vector boson fusion, which are the most relevant channels for the Higgs mass range ${m_h\in[124,127] \GeV}$. We write the normalized Higgs eigenstate $h$ in terms of the gauge components $h_u$, $h_d$ and $s$ as $h=V_{h,h_u} h_u+V_{h,h_d} h_d +V_{h,s} s$. Unless stated otherwise, we consider Higgs production through gluon fusion, which tends to dominate in the case of a sizable $h_u$  admixture in $h$.  The signal strength to diphotons versus the signal strength to dibosons is displayed in Fig.~\ref{fig:ratesZZ}, and versus the signal strength to $b\bar{b}$ in association with a vector boson in Fig.~\ref{fig:ratesbb}. First of all, let us stress that we can always accommodate a signal strength consistent with a SM Higgs in all channels with a total tuning better than 1$\%$. However, it is also possible to explain the anomalous (at the 2$\sigma$-level) signal strength in the diphoton channel measured at both ATLAS and CMS.  As seen in the figures, it is clear that an enhancement in the diphoton signal strength, consistent with the range given in \Eqref{eq:Rgammagamma}, requires at least some reduction in the b-quark signal strength $R_{b\bar{b}}$. This reduction results from a smaller $h_d$ component in $h$ with respect to $h_{SM}$ and thus also leads to a reduction of $R_{\tau\bar{\tau}}$ which is, however, experimentally not very constrained at the moment. More importantly, as can be seen from Figs.~\ref{fig:ratesZZ} and \ref{fig:ratesbb}, the reduction in $R_{b\bar{b}}$ gives rise to an increment in the diboson signal so that $R_{ZZ}\approx R_{\gamma\gamma}$. This could be ruled out in the near future. 

However, in the regions of large total-tuning (red points which also correspond to large values of $\lambda$) one can obtain an enhancement due to charginos \cite{SchmidtHoberg:2012yy} in the diphoton channel while keeping $R_{ZZ}\lesssim R_{\gamma\gamma}$ as shown in Fig.~\ref{fig:enhanced1}.  The enhancement is a consequence of large values of $\lambda$  that lead to a large negative singlet component in $h$, $V_{h,s}\lesssim -0.3$ and a sizable Higgs-chargino coupling. The lightest chargino\footnote{The lightest chargino is not required to be very light due to the large coupling $\lambda$.} mainly Higgsino in composition, can then lead to enhancements in the diphoton channel at loop level that interfere constructively with the $W$-boson contribution
\begin{equation}
R_{\gamma\gamma}\approx \frac{\left|g_{hWW} A^h_1(\tau_{W})+ 4 \frac{m_{W}}{m_{\tilde{\chi}^{\pm}_{1}}} g_{h\tilde{\chi}^{+}_{1}\tilde{\chi}^{-}_{1}}+\dots\right|^2}{\left|A^h_1(\tau_{W})+\frac{4}{3} y_t A^{h}_{1/2}(\tau_t)+\dots\right|^2}~,
\end{equation}
where $\tau_i\equiv m_h^2/(4m_i^2)$, $g_{hWW}$ is the scale-invariant NMSSM Higgs coupling to $W$-bosons and $g_{h\tilde{\chi}^{+}_{1}\tilde{\chi}^{-}_{1}}\approx \sqrt{2} \lambda V_{s,h}$ is the coupling to the lightest chargino pair which being negative and large due to $V_{s,h}$ and $\lambda$,  interferes constructively (though it is a fermion loop) with the $W$-boson contribution. The functions $A^h_1(\tau)$ and $A^h_{1/2}(\tau)$ from loop integration are given by
\begin{equation}
A^h_1(\tau)=-2[2\tau^2+3\tau+3(2\tau-1)f(\tau)]\tau^{-2} \qquad \text{and} \qquad A^h_{1/2}(\tau)=2[\tau+(\tau-1)f(\tau)]\tau^{-2} \, ,
\end{equation}
where $f(\tau)={\rm ArcSin}[\sqrt{\tau}]^2$ for $\tau\leq 1$, and we have used that  $A^h_{1/2}(\tau_{\tilde{\chi}^{\pm}_{1}})\approx 2$. Notice that for $V_{s,h}$ to be significant, one needs large values of $\lambda$ and thus one expects that this type of diphoton enhancement can only be accomplished in the large fine-tuned regions of parameter space. This is what we mostly found and is depicted by the red points with $1.4\lesssim R_{\gamma\gamma}\lesssim 1.5$ shown in \Figref{fig:ratesZZ}. 

However, there are a few additional points with $R_{\gamma\gamma}\gtrsim 1.7$, $1.1\lesssim R_{ZZ}\lesssim 1.32$, $0.35\lesssim R_{b\bar{b}} \lesssim 0.6$ and with total tuning better than 1$\%$ as can be seen in Figs.~\ref{fig:ratesZZ} and \ref{fig:ratesbb}. Most of these points are characterized by having large values of $\tan\beta\gtrsim 4$ and intermediate values of $\lambda\sim 1.2$ such that their corresponding tree-level Higgs masses are close to 126 GeV and therefore their combined tuning is reduced as discussed at the end of \Secref{sec:collider}.  More importantly, the lightest charginos associated with this region are in the range $m_{\tilde{\chi}^{\pm}_{1}}\in [110,120]$ GeV  and $V_{h,s}\lesssim -0.5$ due to a reduction on $V_{h,h_d}$ for large $\tan\beta$, providing a large loop contribution to the diphoton decay. It is interesting that as $\tan\beta$ becomes large, the $h_d$ component of the Higgs diminishes in magnitude, allowing for a large singlet component and moreover, due to the $T$-parameter constraints, we are led to regions of moderate values of $\lambda$ where the total tuning is reduced. So what we learn from the enhanced diphoton region is that a sizable singlet component in the Higgs can be obtained either by:
\begin{enumerate}

\item Large values of $\lambda$ and thus a tuned region (upper-right red corner with $R_{\gamma\gamma}\sim 1.4$ in Fig.~\ref{fig:ratesZZ}).  However this also leads to slightly heavy charginos $m_{\tilde{\chi}^{\pm}_1}\sim \mu$ making the diphoton enhancement mild.

\item Large values of $\tan\beta$ which due to $T$-parameter constraints imply also intermediate values of $\lambda$ and thus smaller tuning (points with $R_{\gamma\gamma}\gtrsim 1.6$ in  Fig.~\ref{fig:ratesZZ}). Since $\lambda$ is not as large one also expects that charginos can be relatively light thus providing a larger loop contribution to the diphoton decay.

\end{enumerate}
Furthermore most of these points correspond to a spectrum where $\tilde{\chi}^0_1$ is either purely singlino or Higgsino with $m_{\tilde{\chi}^0_1}\in[70,120] \GeV$, the $\tilde{\chi}^{\pm}_1$ is purely Higgsino with $m_{\tilde{\chi}^{\pm}_{1}}\in [110,120] \GeV$  and   $\tilde{\chi}^0_2$ is either purely Higgsino or singlino with $m_{\tilde{\chi}^0_2}\in[180,450]$ GeV. Thus the neutralino-chargino constraints discussed in \Secref{sec:collider} do not apply since they are based on the assumption of a mostly electroweak-gaugino  neutralino-chargino sector that is light.  These results suggest that there is region of parameter space with a small amount of tuning and enhanced diphoton decay which deserves further study.

The large singlet component also modifies the coupling to $Z$ and $W$-bosons and down-type quarks, but in the region of enhancement the down-type coupling is further suppressed than the $W$,$Z$-boson coupling, making the production almost close to SM-like.\footnote{The increase in the singlet component of $h$ comes with a decrease in $h_d$, while the $h_u$ component is not much suppressed.} This raises all signal strengths due to the small $BR(h\to b\bar{b})$ but the large singlet component allows a milder enhancement in the $ZZ$ coupling compared to the diphoton coupling (which is more enhanced due to the chargino loop). This is shown in Fig.~\ref{fig:enhanced2} where the ratios of $Z$-boson and diphoton couplings to the SM couplings are plotted with the result that $C_{ZZ}\lesssim C_{\gamma\gamma}$, where
\begin{equation}
C_{XX}\equiv g_{hXX}/g_{h_{SM}XX}\;.
\end{equation}
Notice that the points corresponding to this region possess a negative value of $C_{d\bar{d}}$ as shown in \Figref{fig:enhanced1}, corresponding to a small but negative $h_d$ component in $h$, $V_{h,h_d}\lesssim 0$. This provides an enhancement in the diphoton channel with down-type fermions running in the loop. However due to the smallness of the bottom-quark Yukawa coupling such an enhancement is insignificant and therefore we believe that the fact that these points correspond to a negative $h_d$ Higgs component is only an artifact of the scanning.
\subsection{SUSY searches at the LHC}
\begin{figure}[tb!]\centering
\begin{subfigure}{0.49\linewidth}
\includegraphics[width=\linewidth]{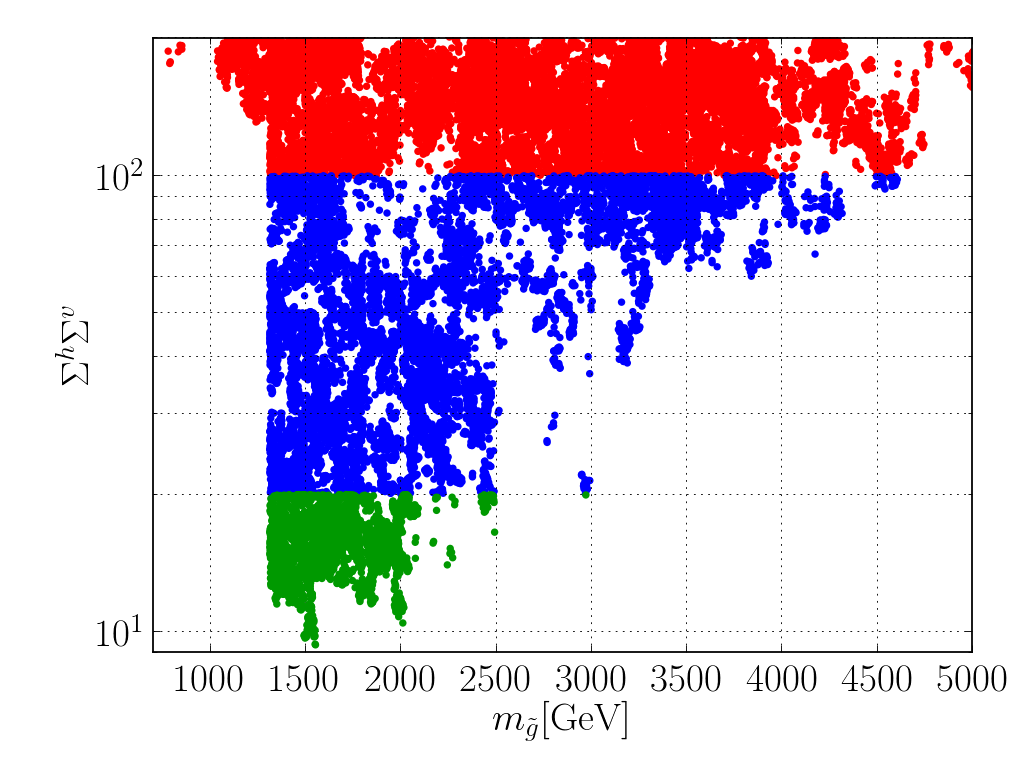}
\caption{\label{fig:Totuningvsmgluino}}
\end{subfigure}
\hfill
\begin{subfigure}{0.49\linewidth}
\includegraphics[width=\linewidth]{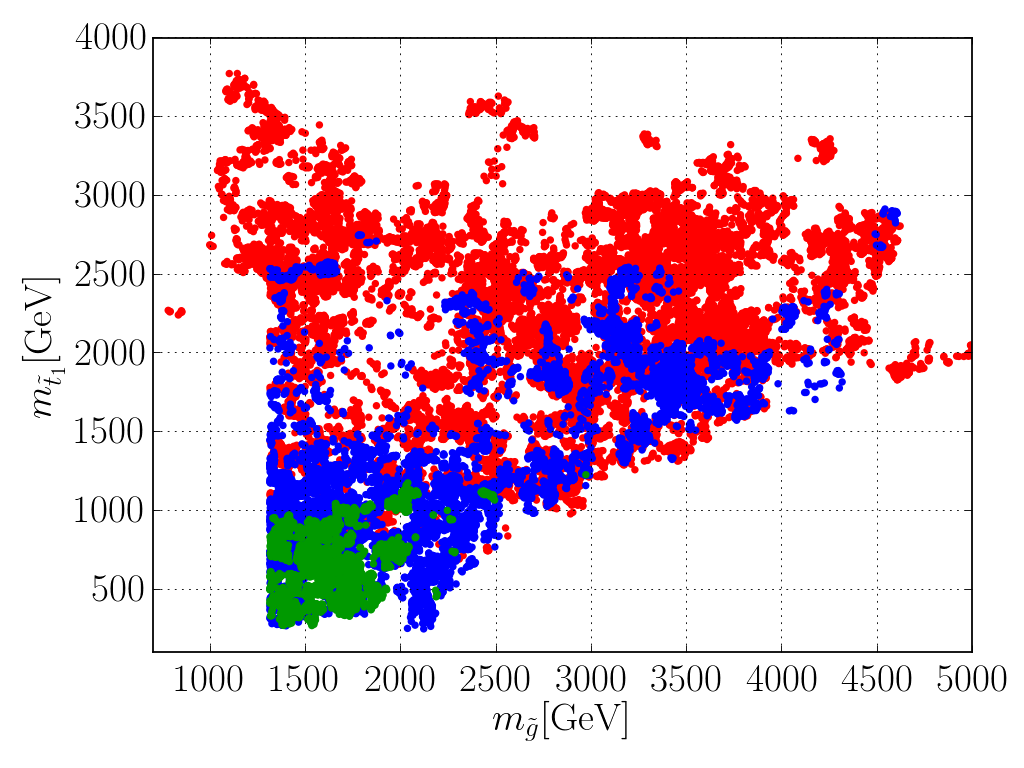}
\caption{\label{fig:stopvsgluino}}
\end{subfigure}
\begin{subfigure}{0.49\linewidth}
\includegraphics[width=\linewidth]{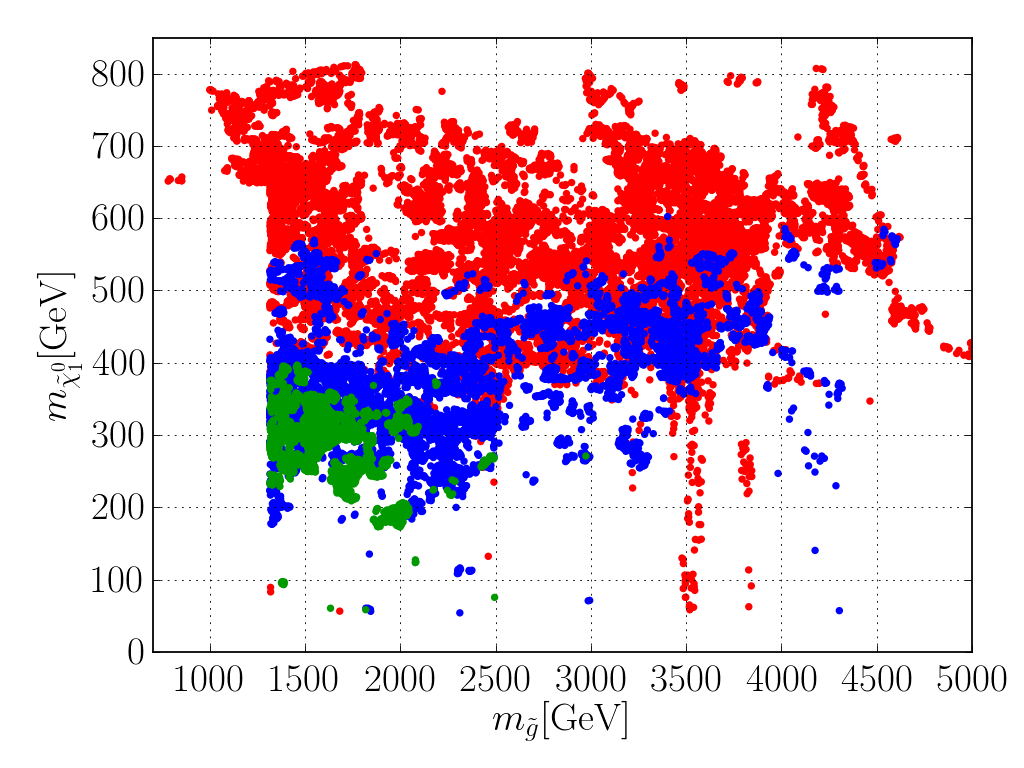}
\caption{\label{fig:mNeutralinovsMgluino}}
\end{subfigure}
\hfill
\begin{subfigure}{0.49\linewidth}
\includegraphics[width=\linewidth]{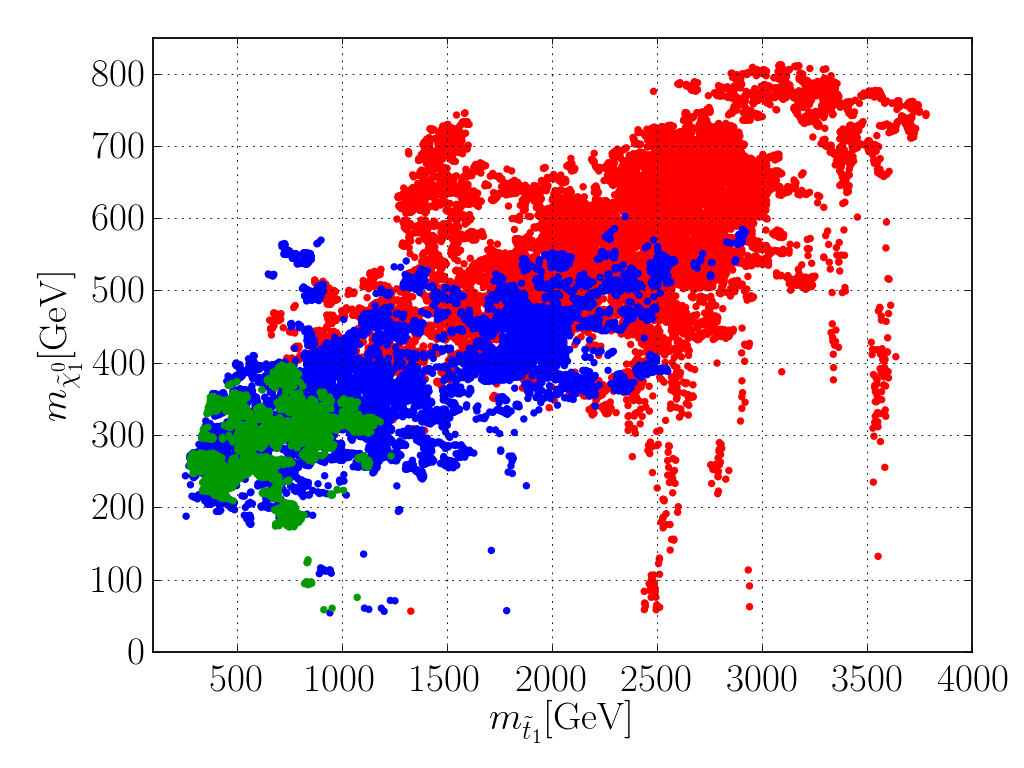}
\caption{\label{fig:mNeutralinovsMstop}}
\end{subfigure}
\caption{\em Scatter plots of (a) the combined tuning $\Sigma^h\Sigma^v$ and (b) the lightest stop mass $m_{\tilde{t}_1}$, as a function of the gluino mass $m_{\tilde{g}}$. 
The scatter plots in (c) and (d) show the lightest neutralino mass $m_{\tilde\chi^0_1}$ as a function of (c) the gluino mass $m_{\tilde g}$ and (d) the lightest stop mass $m_{\tilde t_1}$.  
The green, blue and red points correspond to a combined tuning ($\Sigma^h\Sigma^v$) better than $5\%$, between $1\%$ and $5\%$, and worse than $1\%$, respectively. All points satisfy the constraints discussed in \Secref{sec:collider}.}
\label{fig:SUSYpheno}
\end{figure}

The naturalness constraints in the scale-invariant NMSSM have definite implications for SUSY searches at the LHC. 
We use the combined tuning in the Higgs VEV and the Higgs mass to place naturalness bounds on sparticle masses. We focus on the region with combined tuning better than 1\% (points colored in blue and green) and highlight the optimal region in our scan (corresponding to the green points) with a combined tuning better than 5\%. In addition, all points have a Higgs VEV tuning that is better than $5\%$. As can be seen from the spectrum shown in  \Figref{fig:spectrum}, we find points in our scan with a lightest stop mass $m_{\tilde t_1}$ up to $1.2 (2.6)\TeV$ for a total tuning better than 5\%(1\%). The heavier stop can have masses up to $1.8 (3.5) \TeV$, whereas the sbottom masses can reach up to $1.5 (4.3)\TeV$ for the light sbottom and $7.5 (8)\TeV$ for the heavier sbottom. The fact that we do not have points with sbottom masses above $\sim 8 \TeV$ is due to the chosen range of the input parameters in the scan (cf.~table~\ref{tab:InputParams}).

The dependence of the combined tuning on the gluino mass is shown in \Figref{fig:Totuningvsmgluino}. As expected, it grows with larger gluino masses. For a combined tuning better than 5\%(1\%), we find gluino masses up to $3.0 (4.6) \TeV$. The distribution of the lightest stop mass $m_{\tilde{t}_1}$ versus the gluino mass $m_{\tilde g}$ is plotted in \Figref{fig:stopvsgluino}. The lower bound on $m_{\tilde{t}_1}$, which grows with $m_{\tilde g}$, can be understood from the requirement in \Eqref{eq:noTuningInSquarkParameters} that there is no additional tuning in the squark mass parameters (cf.~footnote~\ref{gluinosucks}). 
As expected, the parameter region with small stop and gluino masses is preferred by small fine-tuning. 
In \Figref{fig:mNeutralinovsMgluino} and \Figref{fig:mNeutralinovsMstop}, we show the results of our scan
in the plane $m_{\tilde \chi_1^0}$ versus $m_{\tilde g}$ and versus $m_{\tilde t_1}$ which is how the experimental results for gluino and stop searches are typically presented.
The lower left corner in the plots in \Figref{fig:SUSYpheno} is already excluded by stop and gluino searches at the LHC, which lead to the limits given in Eqs.~\eqref{eq:GluinoBound} and \eqref{eq:StopBound}. However, as stops and gluinos can be quite heavy for a combined tuning better than $1\%$, we will have to wait for the 14 TeV LHC to completely probe the stop and gluino masses in this natural region of parameter space.

The lightest electroweak charginos and neutralinos, on the other hand, are necessarily light because their mass is directly tied to the parameters entering the Higgs mass. In our scan, we find masses for the lightest chargino up to $405 (625) \GeV$ for a combined tuning better than 5\%(1\%), as well as neutralino masses up to $400 (605) \GeV$ for the lightest neutralino and $420 (675)\GeV$ for the next-to-lightest neutralino. Exploring this mass region for the electroweak charginos and neutralinos will constrain naturalness of the NMSSM at the percent level.

Let us finally mention that for the case $\Mmess=100$ TeV ($\Mmess=1000$ TeV) with combined tuning better than 1$\%$, gluinos are restricted to lie below 3(2.2) TeV, the lightest stop must lie below 2.5(1.3) TeV, whereas a light chargino $m_{\tilde{\chi}^{\pm}_1}\gtrsim 110$ GeV remains possible in the spectrum.



\section{Conclusion}
\label{sec:conclusions}

A well-motivated way to accommodate the recent LHC discovery of a Higgs-like resonance at 126~GeV is to extend the minimal supersymmetric standard model with a singlet superfield $S$. 
In this NMSSM, the Higgs-singlet coupling $\lambda$ can increase the Higgs mass already at tree-level. 
In addition, the sensitivity of the electroweak scale with respect to changes in the soft masses is suppressed at large $\lambda$, further relieving the fine-tuning.
This seems to allow the stop and gluino masses to be increased for the same level of tuning as in the MSSM. However, increasing the coupling above $\lambda \gtrsim 1$ typically also causes the Higgs mass to become heavier than 126 GeV. This then requires a further tuning among the various contributions to the Higgs mass to keep the Higgs light. This additional tuning limits raising the stop and gluino masses. On the other hand as $\lambda$ becomes smaller and we approach the MSSM limit, the tuning increases because we need to rely on large stop soft-mass parameters to raise the Higgs mass. Thus the tuning is minimized when $\lambda\sim 1$.

Identifying the new resonance with the lightest $CP$-even state in the (scale-invariant) NMSSM, then leads to 
interesting constraints on the sparticle spectrum. The naturalness constraint, which clearly requires light superpartners, has to confront the experimental searches from the LHC and electroweak precision tests. In a optimal-case scenario, with split families (including decoupled third-generation sleptons) and a low messenger scale of $20\TeV$, we have performed a numerical scan over the parameter space of the (scale-invariant) NMSSM. For a tuning that is better than $5\%(1\%)$, we find the lightest stop mass to be below 1.2(2.6) TeV, the gluino mass to be below 3.0(4.6) TeV and the lightest charginos and neutralinos to be below 400(600)~GeV.
If the messenger scale is increased to 100(1000) TeV, then the total tuning worsens and becomes below the 5\%(2-3\%)  level.

The measured DM relic abundance can be accounted for by a neutralino LSP, which is dominantly bino-like. The dominant annihilation channels are then via the Higgs and the lightest pseudoscalar. In our scan, however, we mostly find Higgsino- and singlino-like neutralino LSPs, because the parameters determining their mass are related to the recent 126 GeV Higgs mass measurement.

We also find a few points that can explain the experimentally observed enhanced diphoton decay rate of the Higgs.
In particular, this can be achieved via a large value of $\lambda$ and consequently a sizable singlet admixture, leading to a suppression of the coupling $h b\bar b$ and an enhancement of the chargino contribution to the $h\to\gamma\gamma$ amplitude. The chargino can be relatively heavy in this region, since its coupling to the Higgs is enhanced by $\lambda$.
The more interesting possibility occurs for larger values of $\tan\beta$, implying intermediate values of $\lambda$ and small fine-tuning. In this region, the lightest chargino has to be light to give a substantial enhancement of $h\to\gamma\gamma$. Indeed, we find chargino masses of $110-120\GeV$, which will be probed experimentally soon. We believe that this region of small fine-tuning and enhanced Higgs diphoton rate deserves further study.

In summary, our analysis shows that in the optimal-case scenario of a low messenger scale (20 TeV) and light colored third-generation sparticles there are still sizable regions in the parameter space of the scale-invariant NMSSM where the tuning is better than 5\%. This still allows for stop, gluino, chargino and neutralino masses above the current LHC limits that will be eagerly searched for at the 14 TeV LHC, thereby probing naturalness down to the percent level.


\section*{Acknowledgements}
We thank Kaustubh Agashe, Yi Cai, Marcela Carena, Maxim Perelstein, Josh Ruderman, Giovanni Villadoro and 
Carlos Wagner for helpful discussions. This work was supported in part by the Australian Research Council.
TG thanks the SITP at Stanford and CERN TH division for hospitality during the completion of this work. BvH and AM thank SLAC for hospitality. This material is based upon work 
supported in part by the National Science Foundation under Grant No.1066293 and the hospitality of the Aspen Center for Physics.

\section*{Appendix}

\appendix

\section{Mass matrices}
\label{appendixMasses}
In this appendix, we collect the mass matrices of stops, sbottoms and the Higgs-singlet sector (except for the $CP$-even states for which the mass matrix has been discussed already in \Secref{sec:NMSSM}). 

In the $CP$-odd sector of the Higgs-singlet scalars, we first decouple the Nambu-Goldstone boson ``eaten" by the massive $Z$-boson. The remaining $CP$-odd state in the Higgs sector is given by the linear combination $A \equiv \sin\beta\, h_{d,I}+\cos\beta\, h_{u,I}$. The mass matrix in the basis $A_{CP-\rm{odd}}=(A,s_I)$ then takes the form
\bea
M^2_{CP-\rm{odd}}&=&\left(
\begin{array}{ccc}
m_A^2  & -3\kappa v\mu+\frac{m_A^2}{\mu}\lambda v\frac{\sin2\beta}{2}   \\
.  & \lambda^2v^2\frac{m_A^2}{\mu^2}\frac{\sin^22\beta}{4}+\frac{3}{2}\kappa\lambda v^2\sin 2\beta-3 a_{\kappa}\frac{\mu}{\lambda}
\end{array}
\right).
\label{CPoddmatrix}
\eea
Diagonalizing the mass matrix, we obtain the mass eigenstates $a_1$ and $a_2$, where we label the states such that their masses satisfy $m_{a_1}<m_{a_2}$. 

Similar to the MSSM, gauginos and Higgsinos mix after electroweak symmetry breaking. In addition, there is a new mixing mass term proportional to $\lambda$ between the singlino and Higgsinos. In the gauge-eigenbasis $\psi^{0}=(\tilde{B},\tilde{W}^{3}, \tilde{H}^{0}_{d},\tilde{H}^{0}_{u},\tilde{S})$, the neutralino mass matrix is then given by
\begin{eqnarray}
M_{\psi^0}&=&\left(
\begin{array}{cccccc}
M_{1} & 0 & -\cos\beta \sin\theta_W m_{Z} & \sin\beta\sin\theta_{W} m_{Z}  & 0\\
.  & M_2 &  \cos\beta \cos\theta_W m_{Z} & -\sin\beta\cos\theta_{W} m_{Z}  & 0\\
. & . & 0 & -\mu & -\lambda v  \sin \beta\\
. & . & . & 0 & -\lambda v  \cos \beta\\
. & . & . & . & 2 \frac{\kappa}{\lambda}\mu
\end{array}
\right).\nonumber\\
\end{eqnarray}

The charginos mix as in the MSSM. In the gauge-eigenbasis $\psi^{\pm}=(\tilde{W}^{+},\tilde{H}^{+}_{u},\tilde{W}^{-},\tilde{H}^{-}_{d})$, the mass matrix is therefore  given by the usual $2\times 2$ block form
\begin{equation}
\begin{array}{c}
M_{\psi^{\pm}}
\end{array}
=
\left(
\begin{array}{ccc}
0  &  X^{T} \\
X   & 0
\end{array}
\right),
 \qquad
\text{where} \quad
 \begin{array}{c}
X
\end{array}
=
\left(
\begin{array}{ccc}
M_2  & \sqrt{2}m_W \sin\beta\\
\sqrt{2} m_W \cos\beta & \mu
\end{array}
\right)
\end{equation}
and the superscript $T$ denotes the transpose.

The stops and sbottoms also mix as in the MSSM. The mass matrix of the stops is given by
\bea
\left(
\begin{array}{ccc}
a_t  & b_t   \\
b_t  & c_t
\end{array}
\right)
&\equiv&
\left(
\begin{array}{ccc}
m^2_{Q_{3}}+m_t^2+\left(\frac{1}{2}-\frac{2}{3}\sin^2\theta_W\right)m_Z^2\cos2\beta  & m_t(A_t-\frac{\mu}{\tan\beta})  \\
.   & m^2_{u_{3}}+m_t^2+\frac{2}{3}\sin^2\theta_W m_Z^2\cos2\beta 
\end{array}
\right),\nonumber\\
\eea
where the $A_t$-term is defined such that $\Delta \mathcal{L}_{soft}=y_t A_t \tilde{\bar{u}}_{3}\tilde{Q}_{3}H_{u}+c.c.$.  Diagonalizing the mass matrix, we obtain the mass eigenstates  $\tilde{t}_1$ and $\tilde{t}_2$  with masses given by
\begin{equation}
m_{\tilde{t}_2,\tilde{t}_1}^2=\frac{1}{2}(a_t+c_t\pm\Delta_t),
\label{eq:mstop}
\end{equation}
where $\Delta_t \equiv \sqrt{(a_t-c_t)^2+4b_t^2}$. Similarly, the mass matrix of the sbottoms is given by 
\bea
\left(
\begin{array}{ccc}
a_b  & b_b   \\
b_b  & c_b
\end{array}
\right)
&\equiv&
\left(
\begin{array}{ccc}
m^2_{Q_{3}}+m_b^2+\left(-\frac{1}{2}+\frac{1}{3}\sin^2\theta_W\right)m_Z^2\cos2\beta  & m_b(A_b- \mu \tan\beta)  \\
 .    & m^2_{d_{3}}+m_b^2-\frac{1}{3}\sin^2\theta_W m_Z^2\cos2\beta 
\end{array}
\right),\nonumber\\
\eea
and the masses of the mass eigenstates $\tilde{b}_1$ and $\tilde{b}_2$ follow from Eq.~\eqref{eq:mstop} with the replacement $(a_t, b_t, c_t) \rightarrow(a_b, b_b, c_b)$. 

\section{Effective Potential}
\label{appendixPotential}
In this appendix, we give the dominant contributions to the Coleman-Weinberg potential, due to third-generation (s)quarks and the Higgs-singlet sector itself. 
Defining the function 
\begin{equation}
f(m) \equiv m^4 \left(\ln \frac{m^2}{\msoft^2}-\frac32\right)\,,
\end{equation}
we can express these contributions in a compact form.

The contribution due to third-generation (s)quarks to the effective potential is given by
\begin{equation}
V_{1, {\rm Q}} =\frac{N_c}{64\pi^2}\Bigg[2\, \sum_{i=1}^{2} f(m_{\tilde t_i})
+ 2\,\sum_{i=1}^{2} f(m_{\tilde b_i})-4\, f(m_{t})-4\, f(m_{b})\Bigg]\,,
\end{equation}
where $N_c=3$ and $m_{\tilde t_i}$ denotes the stop masses (given in \Eqref{eq:mstop}) and $m_{\tilde b_i}$ the sbottom masses. 
The top and bottom mass are given by $m_t=y_t v_u$ and $m_b=y_b v_d$, respectively.
All masses are evaluated at the renormalization scale $\msoft=\sqrt{m_{Q_3} m_{u_3}}$. The different prefactors (2 for sfermions and -4 for fermions) originate from the degrees of freedom and the different statistics.

The contribution due to the Higgs-singlet sector (including superpartners) is given by
\begin{equation}
 V_{1, {\rm Higgs}}=\frac{1}{64\pi^2}\Bigg[\sum_{i=1}^{3} f(m_{s_i}) + \sum_{i=1}^{2} f(m_{a_i}) -2\,\sum_{i=1}^{5}f(m_{\tilde\chi_i^0})-4\,\sum_{i=1}^{2}f(m_{\tilde\chi_i^\pm})\Bigg]\,.
\end{equation}
The masses are functions of the parameters in the Higgs sector, especially of $\lambda$ and $\kappa$, which can become sizable. An explicit calculation of this contribution to the  effective potential is tedious, since it involves the diagonalization of a $3\times3$ and a $5\times5$ mass matrix. 

The calculation of the minimization conditions and the fine-tuning measure requires the knowledge of several derivatives of the mass eigenvalues with respect to the VEVs as well as parameters in the Lagrangian.
The existence of an analytic expression for the masses is not necessary, however, since an expression for the derivatives can be obtained by differentiating the characteristic polynomial and solving the resulting linear equation for the derivative. This is similar to the calculation of the fine-tuning measure discussed in \Secref{naturalness}.


\bibliography{SNH}

\end{document}